\documentclass[11pt, a4paper, final]{article} 
\usepackage{jheppub}

\title{Deforming and dissecting AdS$_3$ with matter}

\author[1]{Nele Callebaut,}
\author[1]{Blanca Hergueta,}
\author[2]{Ruben Monten}
\author[1]{and Matteo Selle}

\affiliation[1]{Institute for Theoretical Physics, University of Cologne, Zülpicher Str. 77a, 50937 Cologne, Germany}
\affiliation[2]{Theoretical Physics Department, CERN, 1211 Geneva 23, Switzerland}

\emailAdd{nele.callebaut@thp.uni-koeln.de}
\emailAdd{hergueta@thp.uni-koeln.de}
\emailAdd{mselle@thp.uni-koeln.de}
\emailAdd{ruben.monten@cern.ch}

\abstract{
We study deformations of the model by Henneaux, Martínez, Troncoso and Zanelli \cite{Henneaux:2002wm} which features asymptotically AdS$_3$ black hole solutions that incorporate the exact backreaction of a scalar field. 
The presence of bulk matter causes the $T \overline T$ deformation of the (putative) dual CFT$_2$ to differ from the deformation defined in the bulk by imposing Dirichlet boundary conditions at finite radius.
We work out both of these deformations explicitly and verify that $T \overline T$-deforming the boundary theory corresponds to imposing mixed boundary conditions on the metric at the conformal boundary, whereas the bulk “Dirichlet deformation” gives rise to a field theory deforming operator that includes $T \overline T$ as well as other irrelevant terms. 
We check our results by calculating the deformed energy spectrum for either case using both the bulk and boundary prescriptions, finding agreement after taking into account additional terms coming from the flow of the scalar source.
We interpret our explicit results and compare them with the predictions of similar proposals in the literature.
}

\preprint{CERN-TH-2025-268}

\usepackage{amsmath, braket, cleveref, graphicx, tcolorbox, xspace}
\usepackage[dvipsnames]{xcolor}
\newtcolorbox{grayBox}{frame empty, arc=0pt}

\newcommand{\pheq}{\mathrel{\phantom{=}}}
\newcommand{\pd}{\partial}
\newcommand{\td}{\text{d}}

\newcommand{\TT}{\texorpdfstring{\ensuremath{{T \overline T}}}{TTbar}\xspace}

\newcommand{\g}{\gamma}

\renewcommand{\d}{\delta}
\newcommand{\D}{\Delta}
\newcommand{\e}{\epsilon}

\newcommand{\s}{\sigma}

\newcommand{\cO}{\mathcal O}
\newcommand{\cL}{\mathcal L}
\newcommand{\cD}{\mathcal D}
\newcommand{\cW}{\mathcal W}

\newcommand{\Th}{{\hat T}}
\newcommand{\nut}{{\bar{\nu}}}
\newcommand{\st}{{\tilde \s}}
\newcommand{\tti}{{\tilde t}}

\DeclareMathOperator\arctanh{arctanh}
\newcommand{\varphit}{{\tilde \varphi}}
\newcommand{\Bt}{{\tilde B}}
\newcommand{\cLt}{{\tilde \cL}}
\newcommand{\gti}{{\tilde \g}}
\newcommand{\rhot}{{\tilde \rho}}

\newcommand{\mut}{{\tilde \mu}}
\usepackage{tikz}
\usetikzlibrary{calc}
\newcommand{\rt}{{\tilde r}}

\usepackage{ifdraft}
\ifoptiondraft{%
	\usepackage[inline]{showlabels}
    
    \usepackage{fancyhdr}
    \setlength{\headheight}{14.5pt}
    \usepackage{datetime2}
    \fancyhead{} 
    \fancyhead[C]{\textbf{DRAFT} (\DTMnow)}

    \hypersetup{final}
}{}
\usepackage[color=orange!40, obeyDraft, textsize=tiny]{todonotes}
\begin{document}

\maketitle

\section{Introduction}

\subsection{Context and motivation}

    The \TT deformation \cite{Zamolodchikov:2004ce,Smirnov:2016lqw,Cavaglia:2016oda} of a 2d holographic CFT has been conjectured to give access to the dual bulk theory in finite volume \cite{McGough:2016lol}, suggesting a completely new strategy to answer a question which is almost as old as AdS/CFT itself and which was beforehand only addressed in terms of the renormalization group flow \cite{Akhmedov:1998vf,deBoer:1999tgo,Heemskerk:2010hk,Faulkner:2010jy,Balasubramanian:2012hb,Lee:2013dln}.%
    \footnote{Of course, this idea also underlies the formalism of holographic renormalization \cite{Henningson:1998gx,Balasubramanian:1999re,deHaro:2000vlm,Skenderis:1999nb,Skenderis:2002wp,Freidel:2008sh} which has been thoroughly confirmed.} 
    The solvability of \TT-deformed observables and its good UV properties, 
    most evident from the S-matrix \cite{Dubovsky:2012wk, Dubovsky_2013, Dubovsky:2017cnj}, 
    have made this a promising avenue --- perhaps quite surprisingly, in light of the problems associated with 
    imposing different 
    kinds of boundary conditions in gravity \cite{Avramidi:1996ae,Avramidi:1997sh,Anderson:2006lqb,An:2021fcq,Witten:2022xxp,Anninos:2024xhc,Liu:2024ymn,Banihashemi:2024yye,Coleman:2020jte}.
    Many of these problems are avoided by the absence of bulk gravitons in 3d, but it remains unclear how to overcome some other pathologies of \TT-deformed theories with the “holographic” sign in finite volume.
    
    What is clear, is that the presence of matter fields in the bulk forms an objection to the holographic $T\bar T$ conjecture \cite{Kraus:2018xrn,Hartman:2018tkw,Guica:2019nzm, Caputa:2020lpa,Belin:2020oib}. 
    The most direct way to see this, as we will review in this paper, is in the large central charge limit, where the bulk theory is approximately described by a 
    gravitational field theory. 
    The \TT deformation of the CFT partition function can be related to the radial Hamiltonian constraint satisfied by the on-shell action of pure gravity.
    Adding matter to the bulk theory alters the latter in a way that can be mimicked by adding other operators to the field theory deformation. 
    This reasoning 
    follows a reversed direction
    compared to the pure gravity holographic \TT literature: rather than taking the field theory deformation as the starting point and investigating its bulk consequences, it aims to \emph{derive} the field theory deformation starting from the (classical) bulk procedure of restricting to a finite volume and imposing Dirichlet boundary conditions.
    Except in a few special cases, such as bulk conserved currents or sufficiently light scalar fields, these additional terms are irrelevant and do not share the solvability properties of \TT.
    It then seems that the resulting theory must be considered at most as an EFT \cite{Hartman:2018tkw}.
    
    Even within the context of 3d bulk gravity, there are only a number of examples where the classical bulk solution is exactly known, including the backreaction of the matter fields, beyond perturbation theory in Newton's constant.
    In this work, we consider one such model, first explored by Henneaux, Martínez, Troncoso and Zanelli (HMTZ) \cite{Henneaux:2002wm}, which admits asymptotically AdS$_3$ black hole solutions with a nontrivial scalar profile. 
    We assume there exists a UV-completion of this gravitational system for which these solutions are still the saddle points, and which is holographically dual to a CFT$_2$.
    As we will see, this CFT must have a scalar operator $\cO$ of dimension 
    $3/2$ and a stress tensor $T^{ij}$ with a large central charge $c$.%
    \footnote{Here we assumed the “standard quantization” where 
    the field theory source is identified with the leading component in the asymptotic expansion of the scalar field.}
    Although we have no direct access to this putative field theory, we will use the bulk to calculate some of its quantities in the large-$c$ approximation, such as the (holographically renormalized) one-point functions of the stress tensor and scalar operator.
    
    We then consider two ways to deform the field theory.
    The first is the usual \TT deformation.
    We review its holographic dual with asymptotic mixed boundary conditions 
    on the metric and check that the deformed energy levels calculated in this deformed gravitational system are indeed those expected from \TT. 
    The second deformation we consider is instead defined from the bulk perspective: on a finite-sized timelike surface in the bulk which is homologous 
    to the boundary, we consider the induced metric and scalar field, 
    rescaled in such a way that they have a finite value as the surface is taken to the conformal boundary of AdS.
    We then impose Dirichlet boundary conditions and calculate the on-shell value of the (appropriately defined) action, which is an integral over the interior of that surface.
    The deformation is parameterized by the chosen radial coordinate in the bulk.%
    \footnote{As we will see, it is appropriately covariant under reparametrizations of this coordinate.}  
    We derive a flow equation for the on-shell action and find that it contains a \TT component (although not with coefficient one) as well as additional, irrelevant contributions from the scalar operator and source.
    One can check very explicitly that this operator coincides with the generator of Weyl transformations, consistent with \cite{Araujo-Regado:2022gvw}.
    In the undeformed limit, this reproduces the anomaly of the CFT. 
    We furthermore show that the energy levels satisfy \emph{the same} flow equation (up to a sign) \emph{only if} the appropriate change of the scalar source is taken into account.
    This is compatible with field theory expectations, so we take this as a non-trivial check that the deformation we defined from the bulk perspective is indeed consistent with a field theory interpretation.

    In the remainder of this section, we summarize our results.

\subsection{Summary}

    In this paper we consider two distinct deformations of the HMTZ system and its (putative) CFT dual.
    On the field theory side, we formulate these deformations as two one-parameter families of generating functionals $W^{[\mu]}$ and $W^{[\rhot_0]}$ for the \TT and the Dirichlet deformation, respectively.\footnote{Here, $W$ is the generating functional of connected correlation functions. It is given by $-i$ times the logarithm of the `partition function' $Z$ (even though we work in Lorentzian signature, we will still refer to the field theory path integral as the partition function).}
    These generating functionals depend on the field theory metric $\g_{ij}$ and the source $J$ for the scalar operator $\cO$.
    
    The relation between the generating functionals at infinitesimally different values of the parameter is given by an operator constructed from the stress tensor $T^{ij}$ and the scalar operator $\cO$ as well as their background sources, the metric $\g_{ij}$ and scalar source $J$.
    These are holographically related to the (properly renormalized) Brown--York stress  tensor $T_\text{BY}^{ij}$ and the radial momentum $\pi_\phi$ of the scalar, as well as the induced metric $h_{ij}$ and scalar field value 
    $\phi$, respectively.
    As we will see, it will be crucial to keep track of how the sources change (if at all) along the flow, as this changes how the generating functional gets deformed.
    In other words, starting with the generating functional $W_{[0]}[\g, J]$ dual to the HMTZ solution, we will be led to think about different flows in the parameter space spanned by $\mu$, $\g_{ij}$ and $J$.
    
    For the \TT deformation, the deforming operator at constant values of the sources is nothing but 
    \begin{align}
    \label{eq:introWmu}
        \pd_\mu W_{[\mu]}[\g, J] &= \frac12 \int \td^2 x \sqrt{-\g} \braket{\cO_\TT^{[\mu]}}
        \ , &
        \cO_\TT &\equiv T^{ij} T_{ij} - T^2
        \ ,
    \end{align}
    where $T \equiv \g_{ij} T^{ij}$ is the (generically non-vanishing) trace of the stress tensor.
    Its expectation value and that of the scalar operator are, for each $\mu$,
    \begin{align}
        \braket{T_{[\mu]}^{ij}} &= \frac{2}{\sqrt{-\g}} \frac{\d W_{[\mu]}[\g, J]}{\d \g_{ij}}
        \ , &
        \braket{\cO_{[\mu]}} &= \frac1{\sqrt{-\g}} \frac{\d W_{[\mu]}[\g, J]}{\d J}
        \ .
    \end{align}
    As we will review, there exists a simple relation \eqref{eq:TTDeformationJ} between the deformed generating functional $W_{[\mu]}[\g, J]$ and an undeformed generating functional on a different background geometry $W_{[0]}[\g_{[0]}, J]$, where $\g_{ij}^{[0]}$ is 
    a 
    combination of the metric $\g_{ij}$ and stress tensor $T^{ij}$, given in \eqref{eq:deformedMetricAndStressTensor}.
    This relation is schematically represented in \Cref{fig:flows}.
    Applying the GKP/W dictionary to $W_{[0]}[\g_{[0]}, J]$ (note the distinction with the original $W_{[0]}[\g, J]$) we find that it is dual to a bulk geometry where the leading component $\gti_{ij}^{(0)}$ 
    of the Fefferman--Graham expansion of the bulk metric gets identified with $\g_{ij}^{[0]}$. 
    The relation \eqref{eq:TTDeformationJ} then leads to a holographic dual to the \TT-deformed theory with mixed boundary conditions \cite{Guica:2019nzm}: the bulk prescription is that not the asymptotic metric $\tilde{\g}_{ij}^{(0)}$ is kept fixed on the boundary, but instead the combination 
    \begin{align}
    \label{eq:g2g0}
        \g_{ij} &= \gti_{ij}^{(0)} - 2\mu \Th^{\text{BY}}_{ij} + \mu^2 \Th^{\text{BY}}_{ik} \Th^{\text{BY}}_{jl}\gti_{(0)}^{kl}
        \ ,
    \end{align} 
    where $\Th_{ij}^\text{BY} \equiv T_{ij}^\text{BY} - h_{ij} h^{ab}T^\text{BY}_{ab}$ is the trace reversed Brown--York stress tensor evaluated at the asymptotic boundary.
    The boundary conditions for the scalar field (fixing $J$, the rescaled version of $\phi$) do not change.
    We refer to these asymptotic conditions for the metric 
    as “mixed boundary conditions” (MBC):
    \begin{align}
        \text{\underline{MBC}\,:} \qquad \delta \left( \tilde{\g}_{ij}^{(0)} - 2\mu \Th_{ ij}^{\text{BY}} + \mu^2 \Th_{ik}^{\text{BY}} \Th_{jl}^{\text{BY}} \tilde{\g}_{(0)}^{kl} \right) = 0 
        \ .
    \end{align}
    The correct bulk action compatible with these boundary conditions is
    \begin{align}
        S^{(ren)\star}[\gti^{(0)}, \phi^{(0)}] - \mu \int\limits_{\rho \to 0} \td^2 x \sqrt{- \gti_{(0)}} T_{ij}^\text{BY} T_{kl}^\text{BY} \gti_{(0)}^{i[k} \gti_{(0)}^{j]l}
        \ ,
    \end{align}
    where $S^{(ren)*}$ is the on-shell action of the undeformed HMTZ system and $\gti_{ij}^{(0)}$ is given by inverting \cref{eq:g2g0}. 
    A schematic illustration of the MBC duality is given in \Cref{fig:DBCMBCFig}.

    For the Dirichlet deformation, on the other hand, we find that the generating functional gets deformed as 
    \begin{align}
    \label{eq:introWrhot0}
        &\pd_{\rhot_0} W_{[\rhot_0]}[\g, J]
        \\
        &= -\frac12 \int d^2 x \sqrt{-\g} \, f(J) \, \Bigg\langle 
        \begin{aligned}[t]
            &\frac{6\pi}c \cO_\TT+ \frac{3\pi \cO^2}{4c \sqrt{\rhot_0}} + \frac1{2 \rhot_0} \left(\frac{c_2}{2} + c_4 \sqrt{\rhot_0} J^2 - \frac{1}{f(J)} \right) J \cO
            \\
            & + \frac{2c}{3 \pi \rhot_0} \Bigg( \frac{R[\g]}{16} - \frac12 \sqrt{\rhot_0} \g^{ij}\nabla_i J \nabla_j J - \frac1{\rhot_0} \bar{V}_\nu \Bigg) \Bigg\rangle_{[\tilde{\rho}_0]}
            \ ,
        \end{aligned}
        \nonumber
    \end{align}
    where $f(J) \equiv \big( 1 + c_2 \sqrt{\rhot_0}J ^2 + c_4 \rhot_0 J^4 \big)^{-1}$, $\bar V_\nu$ is a renormalized potential 
    and $c_2, c_4$ are  
    holographic renormalization constants.
    We derive this as the boundary consequence of the following bulk definition:
    given an asymptotically AdS$_3$ metric 
    \begin{align}
        \td s^2 &= N^2(\rhot) d\rhot^2 + h_{ij}(\rhot, x) dx^i dx^j 
        \ ,
    \end{align}
    which we refer to as an “asymptotically Fefferman--Graham” coordinate system if $N(\rhot) = 1 / (2\rhot) + \ldots$ 
    at leading order as $\rhot \to 0$, we consider the part $\rhot > \rhot_0$ of spacetime, for some positive $\rhot_0$, and impose boundary conditions where $h_{ij}(\rhot_0)$ and $\phi(\rhot_0)$ are fixed.
    We will call these the (finite, rather than asymptotic) “Dirichlet boundary conditions” (DBC): 
    \begin{align}
        \text{\underline{DBC}\,:} \qquad \delta h_{ij}(\rhot_0) = 0 \ ,  \qquad \d \phi(\rhot_0) = 0 \ .
    \end{align}
    The gravitational boundary term of the action that is compatible with these boundary conditions is the Gibbons--Hawking one at $\rhot = \rhot_0$.
    The field theory metric and source are related to the fixed bulk quantities as $\g_{ij}^{[\rhot_0]} = \rhot_0 \, h_{ij}(\rhot_0)$ and $J^{[\rhot_0]} = \rhot_0^{-1/4} \phi(\rhot_0)$.
    For continuity in the undeformed $\rhot_0 \to 0$ limit, we also keep the appropriate 
    counterterm contributions to the boundary Lagrangian unchanged.
    The duality with DBC is illustrated in \Cref{fig:DBCDBCFig}.

    Comparing \cref{eq:introWmu,eq:introWrhot0}, we see that the deformations indeed coincide in the absence of matter (and using 
    $R[\g] = 0$) if we make the identification $\mu = - 6\pi \rhot_0 / c$. 
    More generally, though, \cref{eq:introWrhot0} contains many additional terms, including a contribution from the irrelevant operator $\cO^2$, which appears to spoil the UV behavior of the deformed theory.

    To check our results, 
    we calculate how the energy of states changes under either deformation, both in the bulk and on the boundary.
    The \TT-deformation in either perspective leads to the well-known 
    square root formula (here without angular momentum and with unit circumference)
    \begin{align}
        E_{[\mu]}=\frac{1}{2\mu}\left(\sqrt{1+4 \mu E_{[0]}}-1\right).
    \end{align}
    The Dirichlet-deformed energy spectrum can be calculated from the bulk perspective to give
    \begin{align}
    \label{eq:introE}
        E_{[\tilde{\rho}_0]}
        &= \frac{c}{12\pi\tilde{\rho}_0} \left( 1 - \sqrt{\rhot_0 F} \, \frac{H + 2B}{H + B} + c_2 \phi^2 + c_4 \phi^4 \right) \ ,
    \end{align}
    where $\phi$, $H$ and $F$ are given in \cref{scalarprofile,F} and where the expression on the right hand side is evaluated at $r \to 1 / \sqrt{\rhot_0}$ and $B = \sqrt{12 \pi E_{[0]} / ((c_4 - 2/3 + 3 \bar \nu / 2) c)}$. 
    From the field theory point of view, a classical argument would suggest that this energy satisfies the same differential equation, up to a sign, as the action \cite{Kruthoff:2020hsi}.
    Here, however, we must be careful not to use the flow equation \eqref{eq:introWrhot0} with the source $J$ kept constant.
    Instead, we must take into account the change of $J^{[\rhot_0]} = \rhot_0^{-1/4} \phi(\rhot_0)$ in the bulk, which leads to a correction term in the differential equation.
    We find that the deformed energy levels \eqref{eq:introE} indeed satisfy the negative of this corrected flow equation, providing evidence that the aforementioned classical relation \cite{Kruthoff:2020hsi} 
    holds holographically at large $c$.

    Besides illustrating the differences between the \TT and the Dirichlet deformation, these results can be compared to the existing literature on (generalizations of) the holographic \TT deformation. 
    We will do so in detail in \cref{sec:discussion} 
    where we e.g.~observe that 
    the presence of a sufficiently light scalar field in the bulk forced us to be careful with the holographic renormalization, giving rise to the factor $f(J)$ in \cref{eq:introWrhot0}.

\subsection{Organization of the paper}
    We begin by reviewing the HMTZ model and solutions of \cite{Henneaux:2002wm} in \cref{sec:HMTZ}.
    We write it in Fefferman--Graham coordinates and holographically renormalize the action.
    In \cref{sec:TTbar} we consider its \TT deformation, reviewing the general dictionary of \cite{Guica:2019nzm} and applying it to the model.
    We confirm that this deformation cannot be interpreted as a finite bulk cutoff and calculate the deformed energies, finding a perfect match with expectations.
    The Dirichlet deformation is defined in \cref{sec:Dirichlet}.
    We derive the deforming operator from the bulk perspective, which equals the generator of Weyl transformations if the field theory metric and sources are kept constant.
    In order to compare the flow of the energies in the bulk and boundary perspectives, we consider instead the deforming operator which takes the bulk flow of the sources into account and find perfect agreement between bulk and boundary flows.
    We end with a comparison of both deformations, a comparison with the literature, and an outlook in \cref{sec:discussion}. In appendix \ref{appendixA} we discuss in detail the boundary conditions for the scalar and the corresponding matter counterterms, generalizing our results for the $T\bar{T}$ deformation to the alternate scalar quantization for the operator of dimension $1/2$.

\section{The HMTZ solution}
\label{sec:HMTZ}

We begin by reviewing the gravitational model considered by Henneaux, Martínez, Troncoso and Zanelli in \cite{Henneaux:2002wm} and its
fully back-reacted solution, which describes a black hole coupled to a scalar field in an asymptotically Anti-de Sitter spacetime. We will assume that there exists a UV completion of this model which has a CFT dual and calculate field theory quantities in the semiclassical approximation by using the AdS/CFT dictionary. The sources for the $\text{CFT}_2$ stress-energy tensor and scalar operator are identified with the corresponding coefficients in the asymptotic expansion of the bulk fields, and we  obtain the semi-classical generating functional of the $\text{CFT}_2$ by holographically renormalizing the gravitational action using the methods 
of \cite{Henningson:1998gx,deHaro:2000vlm,Skenderis:2002wp,deBoer:1999tgo,Papadimitriou:2016yit}. This allows us to compute the renormalized one-point functions of the $\text{CFT}_2$ stress-energy tensor and scalar operator.

\subsection{The gravitational model and its solution}
    HMTZ considered a family of models \cite{Henneaux:2002wm} parameterized by $\nu \ge -1$ with action
    \begin{equation} \label{action}
		S[g,\phi]=\frac{1}{\pi G}\int d^3x \sqrt{-g} \left( \frac{R}{16}-\frac{1}{2}g^{\alpha\beta}\nabla_{\alpha}\phi\nabla_{\beta}\phi-V_{\nu}(\phi) \right) ,
    \end{equation}
    where the potential is given by
    \begin{equation} \label{potential}
		V_{\nu}(\phi)=-\frac{1}{8\ell^2}\big(\cosh^6\phi + \nu \sinh^6\phi\big) 
        = -\frac{1}{8\ell^2}-\frac{3}{8\ell^2}\phi^2 - \frac{\phi^4}{2\ell^2} + O(\phi^6),
    \end{equation}
    with $\ell > 0$ the AdS length. The squared mass of $\phi$ is $m^2 \ell^2 = -3/4$, above the Breiten\-lohner--Freedman bound of $-9/4$.
        
    The model has exact solutions describing a static and circularly symmetric black hole, dressed with a scalar field of the form
    \begin{equation} \label{scalarprofile}
		\phi=\arctanh \sqrt{\frac{B}{H(r)+B}} ,
        \qquad 
        H(r)=\frac{1}{2}\left(r+\sqrt{r^2+4Br}\right) ,
    \end{equation}
    where $B$ is a non-negative integration constant of length dimension one.
    The metric is
    \begin{equation} \label{metriclinelement}
		g_{\alpha\beta}dx^{\alpha}dx^{\beta}=-\left(\frac{H(r)}{H(r)+B}\right)^2F(r)dt^2+\left(\frac{H(r)+B}{H(r)+2B}\right)^2\frac{dr^2}{F(r)}+r^2d\varphi^2,
    \end{equation}
    where 
    \begin{equation} \label{F}
		F(r)=\frac{\big(H(r)\big)^2}{\ell^2} - \nut \left(\frac{3B^2}{\ell^2}+\frac{2B^3}{\ell^2 H(r)}\right) ,
        \qquad \nut \equiv \nu + 1 \ge 0.
    \end{equation} 
    This geometry describes a black hole with mass
    \begin{equation} \label{ESMMBH}
		M_{BH}=\frac{3B^2 \nut}{8G} ,
    \end{equation} 
    and a horizon located at 
    \begin{equation} \label{ESMundeformedrh}
		r_h=B \Theta_{\nu}= 2 B \left(\frac{1}{\sqrt{\nu}}+\sqrt{\nu}\right)\sin \left(\frac{2}{3}\arctan\sqrt{\nu}\right).
    \end{equation}
    Here, $\Theta_{\nu}$ is the first zero of the Schuster function of order $\nu$ \cite{Henneaux:2002wm}. 
    Note that the scalar field \eqref{scalarprofile} is regular everywhere outside the horizon.

    We close this section with the special case $\nut = 0$, for which the metric \eqref{metriclinelement} simplifies to
    \begin{equation} \label{metricnu-1}
		\nut = 0: \;\;\;\;\;\;\;\;\;\;\;\;\;\;\;\;\;\;\;\; g_{\alpha\beta}dx^{\alpha}dx^{\beta}=-\frac{r^2}{\ell^2}dt^2+\frac{\ell^2}{r^2+4Br}dr^2+r^2d\varphi^2.
    \end{equation}
    From here on, we choose units in which $\ell \to 1$.%
    \footnote{In a bit more detail, if we set $B \to \ell B$, $r \to \ell r$ and $t \to \ell t$ then $\ell^2$ just becomes an overall factor in the line element. We can reinstate it by doing the inverse of this transformation.}

\subsection{Holographic analysis} \label{subsection:2.1}

    To straightforwardly identify the putative field theory quantities using the AdS/CFT dictionary, we change coordinates to Fefferman--Graham (FG) gauge with a radial coordinate $\rho$ that vanishes at the asymptotic boundary, 
    \begin{equation} \label{rrhogeneral}
		r=\frac{1}{\sqrt{\rho}}-2B + \left( 1 + \frac34 \nut \right) B^2 \sqrt{\rho} + \frac{4\nut}{3} B^3 \rho +\frac{9\nut}{8}  B^4 \rho^{3/2} + O\big(\rho^2\big) ,
    \end{equation}
    so that the line element \eqref{metriclinelement} is cast into the typical FG form 
    \begin{align}  \label{ESMbulkmetricFGgauge}
		g_{\alpha\beta}dx^{\alpha}dx^{\beta}=&\frac{d\rho^2}{4\rho^2}+\frac{1}{\rho}\Bigg[\gamma_{ij}^{(0)}+\gamma_{ij}^{(1)}\sqrt{\rho}+\gamma_{ij}^{(2)}\rho+\gamma_{ij}^{(3)}\rho^{3/2}+\gamma_{ij}^{(4)}\rho^2+O\left(\rho^{5/2}\right)\Bigg]dx^idx^j
        \nonumber \\
        =&\frac{d\rho^2}{4\rho^2}+\frac{-dt^2+d\varphi^2}{\rho}-4 B \frac{-dt^2+d\varphi^2}{\sqrt{\rho}} 
        + \frac{3 B^2}{2}\Big[-\left( 4 - \nut \right)dt^2 + \left( 4 + \nut \right)d\varphi^2\Big]
        \nonumber
		\\ 
		&-\frac{B^3 \sqrt{\rho}}{3} \Big[ - \left( 12 - 11\nut \right) dt^2 + \left(12 + \nut \right)d\varphi^2\Big]
		\\
		& + \frac{B^4 \rho}{48}\Big[ -\left( 48 - 124 \nut + 27 \nut^2 \right) dt^2 + \left(48 - 76\nut + 27 \nut^2 \right) d\varphi^2 \Big] + O\left(\rho^{3/2}\right),
        \nonumber 
    \end{align} 
    The non-integer powers of $\rho$ are a result of the back-reaction by the scalar field, which itself has the asymptotic expansion 
    \begin{equation} \label{phiexpansion}
		\phi 
        = \phi^{(0)} \rho^{1-\Delta/2} + \dots + \phi^{(2\D - d)} \rho^{\D/2} + \ldots
        = \sqrt{B}\rho^{1/4}+\frac{B^{3/2}}{3}\rho^{3/4}+O\left(\rho^{5/4}\right).
    \end{equation} 
    i.e.~$\Delta = \frac{3}{2}$, $\phi^{(0)} = \sqrt{B}$ and $\phi^{(2\D - d)} = \frac{B^{3/2}}{3}$.%
    \footnote{Here we have chosen $\Delta = 3/2$ rather than the “alternative quantization” $\Delta = 1/2$. The latter would be consistent with imposing asymptotic Neumann boundary conditions on the bulk scalar field, rather than asymptotic Dirichlet for $\Delta = 3/2$, and it would lead to a different field theory dual. 
    We discuss this in appendix \ref{appendixA}.} 
    The scalar also prevents the asymptotic expansion of the metric from truncating at quadratic order in $\rho$, as it does in vacuum AdS$_3$ \cite{Skenderis:1999nb}. 
    Nonetheless, in the special case $\nut=0$, we find that the expansion does truncate. In fact, with the change of coordinates
    \begin{equation} \label{rrhonu-1}
		\nut = 0: \; \; \; \; \; r=\frac{1}{\sqrt{\rho}}-2B+B^2 \sqrt{\rho},
    \end{equation}
    the line element \eqref{metricnu-1} takes the form 
    \begin{equation} \label{gijnu-1}
		\nut = 0: \; \; \; \; \; g_{\alpha\beta}dx^{\alpha}dx^{\beta}=\frac{d\rho^2}{4\rho^2}+  \frac{-dt^2+d\varphi^2}{\rho} \left(1-4B\sqrt{\rho}+6B^2\rho-4B^3\rho^{3/2}+B^4\rho^2\right).
    \end{equation}

    The holographic dictionary in the semiclassical regime $c = 3 / 2G \gg 1$ equates the CFT generating functional $W[\g, J]$, which is a function of the CFT background metric $\g_{ij}$ and the source $J$ for the operator $\cO$, to the gravitational on-shell action as a function of the boundary data $\g_{ij}^{(0)}$ and $\phi^{(0)}$ \cite{Gubser:1998bc,Witten:1998qj} 
    \begin{equation}
    \label{eq:GKP/W}
      W[\g, J] = S^{(ren)\star}[\g^{(0)}, \phi^{(0)}]\Big|_{\g_{ij}^{(0)} = \g_{ij}, \, \phi^{(0)} = J}.
    \end{equation} 
    This requires that the gravitational action has a well-defined variational principle that is consistent with the asymptotic Dirichlet boundary conditions $\d \gamma_{ij}^{(0)} = 0 = \d \phi^{(0)}$ 
    and that it is properly renormalized. 
    We ensure a well-defined variational principle under a Dirichlet condition for the metric at the asymptotic boundary by adding the Gibbons--Hawking boundary term to the action \eqref{action}, which gives what we call the “unrenormalized action”%
    \footnote{Note that the minus sign in front of the GH term is consistent with having the boundary located on the lower bound of the radial coordinate $\rho$. We use the conventions of \cite{Poisson:2009pwt} for the normal vectors, with an inward-pointing normal along the direction of increasing $\rho$.} 
    \begin{align} \label{actionwithb}
		S_{[\e]}^{(un)}[g,\phi] &= S_{[\epsilon]}[g,\phi]+S_{GH}^{[\epsilon]}[h] \nonumber
		\\
		&=\frac{1}{\pi G}\int\limits_{\rho > \e} d^3x \sqrt{-g}\left[\frac{R}{16}-\frac{1}{2}g^{\alpha\beta}\nabla_{\alpha}\phi\nabla_{\beta}\phi-V_{\nu}(\phi)\right]-\frac{1}{8\pi G}\int\limits_{\rho=\e} d^2x \sqrt{-h}K.
    \end{align} 
    Here and in the following, $h$ is the determinant of the induced metric $h_{ij}$ on surfaces $\Sigma_{\rho_0}$ of constant $\rho=\rho_0$, $K=\nabla_{\mu}n^{\mu}$ is the trace of the extrinsic curvature on $\Sigma_{\rho_0}$ and $n^{\mu}$ is the normalized normal vector to $\Sigma_{\rho_0}$. 
    We also introduced a cutoff regulator at $\rho = \e$.

    In order to identify the renormalizing counterterms, we follow \cite{Henningson:1998gx,deHaro:2000vlm,Skenderis:2002wp} and evaluate the divergences of the unrenormalized on-shell action as $\e \to 0$, 
    leading to an expression of the form
    \begin{equation} \label{regonshellaction}
		S^{(un)\star}_{[\epsilon]}[g,\phi] 
        =\int\limits_{\rho=\epsilon} d^2 x \left(a_{(1)}\epsilon^{-1}+a_{(1/2)}\epsilon^{-1/2}+a_{(log)}\log\epsilon\right)+O\left(\epsilon^0\right).
    \end{equation} 
    The renormalizing counterterms are identified so that they cancel these divergences,
    \begin{align} \label{Sctdef}
		S_{[\epsilon]}^{(ct)}[h,\phi]&=S_{ct}^{(1)}[h,\phi]+S_{ct}^{(1/2)}[h,\phi]+S_{ct}^{(log)}[h,\phi]+S_{ct}^{(0)}[h,\phi]+O\left(\sqrt{\epsilon}\right) 
    \end{align} 
    with $S_{ct}^{(n)} = O(\e^{-n})$. 
    The finite counterterm $S_{ct}^{(0)}$ is scheme-dependent.
    In the minimal subtraction scheme it is fixed by the requirement that the terms of order $\e^0$ in $S_{[\epsilon]}^{(ct)}$ vanish, i.e.~$S_{ct}^{(0)}$ is chosen to cancel the (subleading) order $\e^0$ contributions from $S_{ct}^{(n>0)}$. 
    The renormalized action is thus defined as $S^{(ren)}[g,\phi] = \lim\limits_{\epsilon\rightarrow 0}S^{(ren)}_{[\epsilon]}[g,\phi]$ with
    \begin{align}
    \label{Srengeneral}
        S^{(ren)}_{[\epsilon]}[g,\phi] &= S^{(un)}_{[\epsilon]}[g,\phi]+S_{[\epsilon]}^{(ct)}[h,\phi]
        \nonumber \\ 
        &= \frac{1}{\pi G}\int_{\rho > \e} d^3x \sqrt{-g}\left[\frac{R}{16}-\frac{1}{2}g^{\alpha\beta}\nabla_{\alpha}\phi\nabla_{\beta}\phi-V_{\nu}(\phi)\right]
        \\
        &\pheq -\frac{1}{8\pi G}\int\limits_{\rho=\e} d^2 x \sqrt{-h}\big(K+1+ c_2 \phi^2+c_4\phi^4\big),
    \end{align} 
    with $c_2 = 2$ and a scheme-dependent constant $c_4$, which is $2/3$ in the minimal subtraction scheme. 
    These counterterms allow for a well-posed variational principle under an asymptotic Dirichlet condition for the scalar. The detailed derivation of this result and its analog for asymptotic Neumann boundary conditions are reviewed in appendix \ref{appendixA}. 
    The counterterm canceling the logarithmic divergence is related to the conformal anomaly $\mathcal{A}$ as 
    \begin{equation} \label{confandef}
		S_{ct}^{(log)}[h,\phi] \sim \int\limits_{\rho=\epsilon} d^2 x \sqrt{-h} \; \mathcal{A} \log\epsilon \sim  \epsilon \, \partial_{\epsilon}S^{(ren)\star}_{[\epsilon]}[g,\phi]
    \end{equation}
    and for our solution we find $a_{(log)}\sim R\left[\gamma^{(0)}\right]=0$.

    Using the renormalized action, we can calculate the one-point functions 
    \begin{equation} \label{Orendef}
		\braket{\mathcal{O}}
        \equiv \frac{1}{\sqrt{-\g}} \frac{\delta W}{\delta J}
        =\lim_{\e \rightarrow 0}\frac{\e^{-3/4}}{\sqrt{-h}}\frac{\delta S^{(ren)\star}_{[\epsilon]}}{\delta \phi}
    \end{equation} 
    and 
    \begin{equation} \label{Tijdef}
		\braket{T_{ij}}
        = -\frac{2}{\sqrt{-\gamma}}\frac{\delta W}{\delta \gamma^{ij}}=-\lim_{\e \rightarrow 0}\frac{2}{\sqrt{-h}}\frac{\delta S^{(ren)\star}_{[\epsilon]}}{\delta h^{ij}} \equiv T_{ij}^\text{BY},
    \end{equation} 
    where the right-hand side is nothing but the Brown--York stress tensor in the bulk. 
    Using the explicit solution \eqref{ESMbulkmetricFGgauge}-\eqref{phiexpansion} 
    we find 
    \begin{align} \label{generalonepointfcn}
		\braket{\mathcal{O}} &= \left(1-\frac{3}{2}c_4\right)\frac{B^{3/2}}{3\pi G } ,
        \nonumber \\
		\braket{T_{ij}} \td x^i \, \td x^j &= \frac1{8\pi G} \left( \gamma_{ij}^{(2)} - \left( c_4 + \frac{16}3 \right) B^2 \gamma_{ij}^{(0)} \right)  \td x^i \, \td x^j 
        \\
        &= \cL_t \td t^2 + \cL_\varphi \td \varphi^2
        \ , 
    \end{align} 
    where we defined 
    \begin{align}
    \label{eq:LtLphi}
        \cL_t &\equiv \frac{B^2}{8\pi G} \left( c_4 - \frac23 + \frac{3\nut}2 \right)
         \ , &
        \cL_\varphi &\equiv -\frac{B^2}{8\pi G} \left( c_4 - \frac23 - \frac{3\nut}2 \right)
        \ .
    \end{align}
    Independently from the choice of subtraction scheme, the following relation between the conformal anomaly \eqref{confandef} and the one-point functions of the scalar operator and of the trace of the dual stress tensor holds for our solution \cite{Papadimitriou:2016yit}: 
    \begin{equation} \label{anomalyrelationOT}
		 \mathcal{A} \sim \frac{J}{2} \braket{\mathcal{O}} -  \braket{T} =0.
    \end{equation}

\section{\TT deformation} \label{sec:TTbar}

Now that we have identified the basic properties of the holographic dual to the HMTZ system, in this section we set out to deform it using the \TT operator. We are guided here by the field theory and derive the gravitational consequences of the \TT deformation from the GKP/W dictionary. 
Following the variational method of \cite{Bzowski:2018pcy, Guica:2019nzm}, we find that the gravitational dual of the \TT-deformed field theory has mixed boundary conditions for the metric and unchanged asymptotic boundary conditions for the scalar field in the bulk. We construct the appropriate gravitational boundary term ensuring a well-posed variational problem under such boundary conditions and calculate the deformed operator expectation values. 
Unlike in the case of pure gravity, we show that the mixed boundary conditions here are not generally equivalent to Dirichlet boundary conditions at a finite cutoff. Finally, we employ the duality to compute the deformed energy, finding precise agreement with the well-known formula for the \TT-deformed spectrum \cite{Dubovsky:2012wk,Caselle:2013dra,Smirnov:2016lqw, Cavaglia:2016oda}. 
The energy matching indicates that the deformation involves changing how a given boundary state is identified on the gravitational side, leading to a flow of the 
state parameter 
labeling the bulk solutions. 
In appendix \ref{appendixA} we comment on how this duality extends to the alternate choice of 
scalar quantization. 

\subsection{Boundary flow equation for \TT} \label{sec:3.1}
    
    The \TT deformation in the boundary field theory changes its action $S_{[\mu]}$ as 
    \begin{align}
        \frac{d}{d\mu} S_{[\mu]} &= \frac12 \int \td^2 x \sqrt{-\g} \,\cO_\TT^{[\mu]}
        \ , &
        \cO_\TT &= T^{ij} T_{ij} - T^2 = -\e_{ik} \e_{jl} T^{ij} T^{kl} 
        \ ,
    \end{align} 
    where 
    $\g_{ij}$ 
    is the metric of the CFT background geometry, $\e_{ij}$ is the Levi--Civita tensor and 
    $\sqrt{-\g} \, T_{[\mu]}^{ij} = 2 \d S_{[\mu]} / \d \g_{ij}$ 
    is the instantaneous stress tensor of the deformed theory.     
The partition function of the deformed theory at deformation parameter $\mu$ can be written, by a Hubbard--Stratonovich transformation \cite{Cardy:2018sdv,Bzowski:2018pcy,Dubovsky:2018bmo,Guica:2019nzm,Tolley:2019nmm}, in terms of the theory with infinitesimally different value $\mu - \delta \mu$ as 
\begin{align}
    \label{eq:Zmu} 
        Z_{[\mu]}[\g] &= \int \cD \psi \, e^{i S_{[\mu - \d \mu]}[\g] + \frac{i}2 \int \td^2 x \sqrt{-\g} \d \mu_{ij, kl} T_{[\mu - \d \mu]}^{ij} T_{[\mu - \d \mu]}^{kl}} 
        \nonumber \\
        &\pheq \times \sqrt{\det \d \mu^{-1}} \int \cD \st \, e^{-\frac{i}2 \int \td^2 x \sqrt{-\g} \, \st_{ij} (\d \mu^{-1})^{ij, kl} \st_{kl} }
        \ ,
\end{align}
where the matrix $\d \mu_{ij, kl} \equiv -\d \mu \, \e_{ik} \e_{jl}$ has the inverse $(\d \mu^{-1})^{ij, kl} = -(\e^{ik} \e^{jl} + \e^{il} \e^{jk}) / (2\d \mu)$ 
    and the second line of \cref{eq:Zmu} is just one. Shifting $\st_{ij} = \s_{ij} + \d \mu_{ij, kl} T_{[\mu - \d \mu]}^{kl}$, 
    this gives 
\begin{align}
\label{eq:Zmuh2} 
        Z_{[\mu]}[\g] &= \sqrt{\det \d \mu^{-1}} \int \cD \psi \, \cD \s e^{i S_{[\mu - \d \mu]}[\g] - i \int \td^2 x \sqrt{-\g} \s_{ij} T_{[\mu - \d \mu]}^{ij} } e^{-\frac{i}2 \int \td^2 x \sqrt{-\g} \s_{ij} (\d \mu^{-1})^{ij, kl} \s_{kl}} 
\end{align}
as an integral over not just the matter fields $\psi$ of the field theory, but also an auxiliary field $\sigma$ that couples to the stress tensor.     
    If we assume that $\s_{ij}$ is infinitesimal, which we will see is justified a posteriori, we can recognize the first exponential as $e^{i S_{[\mu - \d \mu]}[\g - 2 \s]}$, i.e.~the undeformed action integrated over the manifold with metric $\g_{ij} - 2 \s_{ij}$, so that we can perform the integral over $\psi$ and find, up to higher-order corrections in $\s_{ij}$,
    \begin{align} 
        Z_{[\mu]}[\g] &\approx \sqrt{\det \d \mu^{-1}} \int \cD \s Z_{[\mu - \d \mu]}[\g - 2\s] e^{-\frac{i}2 \int \td^2 x \sqrt{-\g} \s_{ij} (\d \mu^{-1})^{ij, kl} \s_{kl}}
        \ .
    \end{align} 
    We see that the aforementioned assumption is self-consistent using the method of steepest descent, which becomes arbitrarily accurate as $\d \mu \to 0$: the value of $\s_{ij}$ at the saddle point is itself infinitesimal 
    \begin{align}
        \s_{ij}^\star = -\d \mu_{ij, kl} \braket{T_{[\mu - \d \mu]}^{kl}}
        \ .
    \end{align}
    We thus find 
    \begin{align}
    \label{eq:Zmuh2Z0}
        Z_{[\mu]}[\g] &= Z_{[\mu - \d \mu]}[\g_{ij} + 2 \d \mu_{ij, kl} \braket{T_{[\mu - \d \mu]}^{kl}}]  e^{-\frac{i \d \mu}2 \int \td^2 x \sqrt{-\g} \braket{\cO_\TT^{[\mu - \d \mu]}}}
        \ ,
    \end{align} 
    where we used factorization of the $\cO_\TT$ one-point function \cite{Zamolodchikov:2004ce}.
    
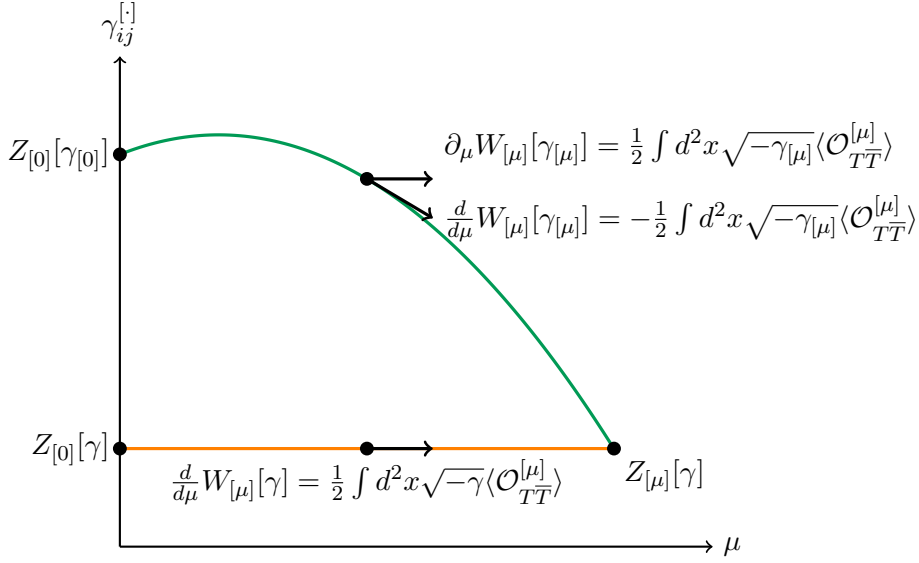
\begin{figure}
    \centering
    \begin{tikzpicture}[scale=1.3]
    
    \draw[->, thick] (0,0) -- (6,0) node[right] {$\mu$};
    \draw[->, thick] (0,0) -- (0,5) node[above] {$\gamma^{[\cdot]}_{ij}$};
    
    \coordinate (Z0gamma) at (0,1);
    \coordinate (Z0gamma0) at (0,4);
    \coordinate (ZmuGamma) at (5,1);
    
    \draw[very thick, orange] (ZmuGamma) -- (Z0gamma);
    
    \draw[very thick, ForestGreen, domain=0.:5., variable=\t, samples=100] plot ({\t},{4 + 2 * \t / 5 - \t * \t / 5});

    \fill (Z0gamma) circle (2pt);
    \fill (Z0gamma0) circle (2pt);
    
    \node[left] at (Z0gamma) {$Z_{[0]}[\gamma]$};
    \node[left] at (Z0gamma0) {$Z_{[0]}[\gamma_{[0]}]$};
    
    \fill (ZmuGamma) circle (2pt);
    \node[below right] at (ZmuGamma) {$Z_{[\mu]}[\gamma]$};
    
    \coordinate (MidStraight) at (2.5, 1);
    \fill (MidStraight) circle (2pt);
    
    \draw[very thick, ->] (MidStraight) -- ++(2/3, 0);
    \draw (MidStraight) node[below]
        {$\frac{d}{d\mu}W_{[\mu]}[\gamma]=\frac{1}{2}\int d^2x \sqrt{-\gamma}\langle \mathcal{O}^{[\mu]}_\TT \rangle$};
    
    \coordinate (MidCurve) at (2.5, 15 / 4);
    \fill (MidCurve) circle (2pt);
    
    \draw[very thick, ->] (MidCurve) -- ++(2/3, 0) node[above right] {$\partial_{\mu}W_{[\mu]}[\gamma_{[\mu]}]= \frac{1}{2}\int d^2 x\sqrt{-\gamma_{[\mu]}}\langle \mathcal{O}^{[\mu]}_\TT \rangle$};
    \draw[very thick, ->] (MidCurve) -- ++(2/3, -2 / 5) node[right] {$\frac{d}{d\mu}W_{[\mu]}[\gamma_{[\mu]}] = - \frac{1}{2}\int d^2x \sqrt{-\gamma_{[\mu]}} \langle \mathcal{O}^{[\mu]}_\TT \rangle$};
    
        
    
    
    \end{tikzpicture}
    \caption{Geometric representation on the $(\mu,\gamma^{[\cdot]}_{ij})$ plane of the auxiliary flow \cref{eq:Zmu2Z0} compared to the standard field theory flow, represented respectively by the curved (green) and straight (orange) trajectories. Along the auxiliary flow, the background geometry $\gamma_{ij}^{[\mu]}$ is non-constant, flowing from $\gamma_{ij}^{[0]}$ to $\gamma_{ij}$ according to $\gamma_{ij}^{[\mu]} = \gamma_{ij}^{[0]} - 2 \mu \braket{\hat T^{[0]}_{ij}} + \mu^2 \braket{\hat T^{[0]}_{ik}} \braket{T^{[0]}_{jl}}  \gamma^{kl}_{[0]}$. The short arrows indicate the corresponding $\mu$-derivatives, illustrating their relation to the $\langle \mathcal{O}_\TT \rangle$ operator.}
\label{fig:flows}
\end{figure}

    In other words, we found that the partition function $Z_{[\mu]}[\g]$ of a 
    \TT-deformed theory on a manifold with metric $\g_{ij}$ is related to the partition function $Z_{[\mu - \d \mu]}[\g_{[\mu - \d \mu]}]$ on the “auxiliary metric” $\g_{ij}^{[\mu - \d \mu]} \equiv \g_{ij} + 2\d \mu \braket{T_{ij}^{[\mu]} - \g_{ij} T_{[\mu]}} + O(\d \mu^2)$.
    This relation is familiar from the relation between the \TT deformation and random geometries \cite{Cardy:2018sdv} and the relation with string theory \cite{Callebaut:2019omt} where a world-sheet and a target-space metric appear. 
We want to emphasize the difference between performing an infinitesimal amount of \TT deformation while keeping the metric fixed, which would compare $Z_{[\mu]}[\g]$ to $Z_{[\mu - \d \mu]}[\g]$, and the relation given in \cref{eq:Zmuh2Z0} between $Z_{[\mu]}[\g]$ to $Z_{[\mu - \d \mu]}[\g_{[\mu - \d \mu]}]$.   
    We will distinguish them by referring to relations such as \cref{eq:Zmuh2Z0}, where the background metric changes as a function of the \TT deformation parameter, as “auxiliary \TT relations”.
    We schematically represent these two ways to relate different partition functions in \Cref{fig:flows}.

    We can now interpret the equation \eqref{eq:Zmuh2Z0} as the result of changing both the auxiliary metric and the partition function, or equivalently the generating functional $W = -i \ln Z$, according to the system of equations 
    \begin{align}
    \label{eq:TTDeformation}
        \frac{\td}{\td \mu} W_{[\mu]}[\g_{[\mu]}] &= -\frac12 \int \td^2 x \sqrt{-\g_{[\mu]}} \braket{\cO_\TT^{[\mu]}}
        \ , 
        \\ 
    \label{eq:metricFlow}
        \pd_\mu \g_{ij}^{[\mu]} &= -2 \langle T_{ij}^{[\mu]} - T_{[\mu]} \g_{ij}^{[\mu]}\rangle
        \ .
    \end{align} 
    We must furthermore demand that \cref{eq:Zmuh2Z0} 
    is compatible with the definition of the stress tensor, which gives
    \begin{equation}
    \label{eq:TmuDef}
        \d W_{[\mu]} = \frac12 \int d^2x \sqrt{-\g_{[\mu]}} \braket{T^{ij}_{[\mu]}} \d \g^{[\mu]}_{ij} \ ,
    \end{equation} 
    and which leads to a derivative equation that involves the stress tensors 
    \begin{align}
    \label{eq:TDeformation}
        \pd_\mu \left( \sqrt{-\g_{[\mu]}} \braket{T_{ij}^{[\mu]}} \right)
        = \sqrt{-\g_{[\mu]}} \Braket{ -\frac12 \g^{[\mu]}_{ij} \cO_\TT^{[\mu]} - 2 T^{[\mu]}_{ik} T^{[\mu]}_{jl} \g_{[\mu]}^{kl} +2 T^{[\mu]}_{ij} T^{[\mu]} }
        \ .
    \end{align}
    It was shown in \cite{Guica:2019nzm} that the system of differential \cref{eq:metricFlow,eq:TDeformation} 
    has a closed form solution, which is conveniently expressed using $\Th_{ij} \equiv T_{ij} - \g_{ij} T_{kl} \g^{kl}$. 
    This solution relates any two points along the “auxiliary” (green) line in \Cref{fig:flows}. In particular it relates $Z_{[\mu]}[\g]$ to an undeformed partition function $Z_{[0]}[\g_{[0]}]$ with auxiliary metric
    \begin{align}
    \label{eq:deformedMetricAndStressTensor}
        \g_{ij}^{[0]} &= \g_{ij} + 2\mu \braket{\Th_{ij}^{[\mu]}} + \mu^2 \braket{\Th_{ik}^{[\mu]}} \braket{\Th_{jl}^{[\mu]}} \g^{kl}
        \ , \nonumber \\
        \text{where} \qquad \braket{\Th_{ij}^{[\mu]}} &= \braket{\Th_{ij}^{[0]}} - \mu \braket{\Th_{ik}^{[0]}} \braket{\Th_{jl}^{[0]}} \g_{[0]}^{kl} 
        \ .
    \end{align}
    We have used that the metric $\g_{ij}^{[\mu]}$ at the end of the flow is the field theory one $\g_{ij}$. 
    Since $\sqrt{-\g_{[0]}} \braket{T_{[0]}^{ij}} = 2 \d W_{[0]} / \d \g_{ij}^{[0]}$, 
    where the undeformed generating functional $W_{[0]}$ is assumed to be known, this implicit equation for $\g_{ij}^{[0]}$ can be solved in terms of $\g_{ij}$. From \cref{eq:deformedMetricAndStressTensor} one can derive 
        \begin{equation}
        \label{eq:gO}
            \sqrt{-\g} \, \braket{\cO^{[\mu]}_\TT} = \sqrt{-\g_{[0]}} \, \braket{\cO_\TT^{[0]}}
            \ .
        \end{equation}
    Given this solution, the full proposal for the semiclassical partition function is 
    \begin{align}
    \label{eq:Zmu2Z0} 
        Z_{[\mu]}[\g] = Z_{[0]}[\g_{[0]}] e^{-\frac{i \mu}2 \int \td^2 x \sqrt{-\g_{[0]}} \braket{\cO_\TT^{[0]}}}
        \ ,
    \end{align} 
    where, for fixed $\g_{ij}$ on the left-hand side, the appropriate pair $\g_{ij}^{[0]}$ and $T_{ij}^{[0]}$ (the latter being derived from $Z_{[0]}$) which solves \cref{eq:deformedMetricAndStressTensor} is used on the right-hand side. Equation \eqref{eq:Zmu2Z0} provides a Legendre transform relation between $W_{[\mu]}[\g]$ and $W_{[0]}[\g_{[0]}]$ \cite{Gubser:2002vv, Witten:2001ua}, with the choice of independent variable ($\gamma$ or $\gamma_{[0]}$) for the description of physics having its analogue in the context of thermodynamics as a choice of ensemble.  

    We reemphasize that this relation between partition functions along the green line in \Cref{fig:flows} is \emph{not} identical to the \TT flow that relates the undeformed field theory to the deformed one \emph{on the same manifold} $\g_{ij}$, along the orange line in \Cref{fig:flows}. 
    For example, \cref{eq:gO} does not imply that the deforming operator $\cO_\TT$ is constant along the usual \TT flow (in orange). We regard the undeformed partition function on the right-hand side of \cref{eq:Zmu2Z0} merely as an auxiliary system that can be used to do computations in the deformed theory. 

    Finally, we should comment on the addition of matter. 
    In the presence of nontrivial sources for CFT operators, such as a source $J\equiv J_{[\mu]}$ for the scalar operator $\cO_{[\mu]}$, the derivation above can be repeated for $Z_{[\mu]}[\g, J]$ 
    in either of two ways, depending on whether one defines the stress tensor to be the variation of $S_{[\mu]}[\g]$ or of $S_{[\mu]}[\g] + \int \td^2 x \sqrt{-\g} \, J_{[\mu]} \cO_{[\mu]}$ under changes of $\g_{ij}$. 
    If we choose the former, we find that the undeformed partition function appearing in \cref{eq:Zmuh2} is not $Z_{[\mu - \d \mu]}[\g - 2\s, J]$ but rather $Z_{[\mu - \d \mu]}[\g - 2\s, J (1 + \g^{ij} \s_{ij})]$, 
    which leads to more complicated differential equations that involve a nontrivial flow of $J$.%
    \footnote{%
        One way out of this issue is to identify the scalar density $\sqrt{-\g} J$,  
        rather than the scalar field $J$, as the source and keep it as an independent variable while varying $\g_{ij}$.
        It would be interesting to formulate the holographic dictionary using this perspective, but we postpone this to future work.
    }
    The other choice, which was used in \cite{Guica:2019nzm}, is instead consistent with $2\d W_{[\mu]}[\g_{[\mu]}, J_{[\mu]}] / \d \g^{[\mu]}_{ij} =\sqrt{-\gamma_{[\mu]}} \braket{T^{ij}_{[\mu]}}$.
    It formally leads to \cref{eq:Zmu2Z0} with $Z_{[\mu]}[\g] \to Z_{[\mu]}[\g, J]$ and $Z_{[0]}[\g_{[0]}] \to Z_{[0]}[\g_{[0]}, J_{[0]}]$.
    However, generic values of $J$ break translational and rotational invariance, and hence alter the stress tensor conservation equations.
    Without conservation, it is not clear that one can properly define the $\cO_\TT$ operator as in \cite{Zamolodchikov:2004ce}.
    In the present case, we nevertheless get away with this dictionary, because we will only encounter static and circularly symmetric sources $J$; indeed, the scalar field $\phi$ only depends on the $\rho$ coordinate. 
    Given the solution \eqref{eq:deformedMetricAndStressTensor} for the auxiliary metric, the natural solutions for the deformed scalar source $J_{[\mu]}$ and corresponding deformed scalar operator $\langle\mathcal{O}_{[\mu]}\rangle=\frac{1}{\sqrt{-\gamma_{[\mu]}}}\frac{\delta W_{[\mu]}}{\delta J_{[\mu]}}$ are%
    \footnote{The variational method requires
		\begin{equation*}
			\sqrt{-\gamma_{[0]}}\; \langle \mathcal{O}_{[0]} \rangle \, \delta J_{[0]} = \sqrt{-\gamma_{[\mu]}}\; \langle \mathcal{O}_{[\mu]}\rangle \, \delta J_{[\mu]} \ .
		\end{equation*}
        Thus, a more general solution is provided by
		\begin{equation*}
			 J_{[\mu]} = F_{[\mu]} \left[ J_{[0]}\right], \qquad \qquad \langle\mathcal{O}_{[\mu]} \rangle= \left( \frac{\delta F_{[\mu]}}{\delta J_{[0]}}\right)^{-1} \sqrt{\frac{\gamma_{[0]}}{\gamma_{[\mu]}}}\;\langle\mathcal{O}_{[0]}\rangle,
	    \end{equation*}
        with $F_{[\mu]}$ a general functional of $J_{[0]}$ reducing to the identity in the undeformed limit, i.e. $F_{[0]}=J_{[0]}$.}
    \begin{equation} \label{defJO}
    J_{[\mu]}=J_{[0]}\equiv J, \;\;\;\;\; \qquad \qquad \langle \mathcal{O}_{[\mu]}\rangle=\frac{\sqrt{-\gamma_{[0]}}}{\sqrt{-\gamma_{[\mu]}}}\langle \mathcal{O}_{[0]}\rangle.
    \end{equation}
    Hence, we will use 
    \begin{align}
    \label{eq:TTDeformationJ}
        Z_{[\mu]}[\g,J] = Z_{[0]}[\g_{[0]},J] e^{-\frac{i \mu}2 \int \td^2 x \sqrt{-\g_{[0]}} \braket{\cO_\TT^{[0]}}}  \, .
    \end{align}

    The generating functional satisfies 
    \begin{align}
    \label{eq:TTpartial}
        \partial_{\mu}W_{[\mu]}[\gamma_{[\mu]},J]= \frac{1}{2}\int d^2x \sqrt{-\gamma_{[\mu]}}\langle\mathcal{O}^{[\mu]}_{T\bar{T}}\rangle \ ,
    \end{align} 
    with a \emph{partial} derivative with respect to the deformation parameter, denoting that the sources are kept constant. 
    The sign in the right hand side coincides with that of the standard \TT flow $\frac{d}{d\mu} W_{[\mu]}[\gamma,J]=\frac{1}{2}\int d^2x \sqrt{-\gamma}\langle\mathcal{O}^{[\mu]}_{T\bar{T}}\rangle$ with total derivative along the orange line in Figure \ref{fig:flows}, i.e.~with fixed $\gamma$.

\subsection{A bulk dual with asymptotic mixed boundary conditions}
\label{sec:TTbulkDer}

    \par We will now review why the generating functional of the deformed theory can be calculated using an \emph{auxiliary bulk solution} for which the boundary metric 
    is given by 
    a mixture of the field theory metric and its stress tensor \eqref{eq:deformedMetricAndStressTensor} \cite{Guica:2019nzm}. We choose to fix the background geometry of the field theory to $\gamma_{ij}dx^idx^j=-dt^2+d\varphi^2$ and we introduce the auxiliary bulk solution $\tilde{g}_{\alpha\beta}dx^{\alpha}dx^{\beta}$. To distinguish it from the bulk geometry that gives $Z_{[0]}[\g]$, we decorate the auxiliary bulk metric and its FG metric coefficients 
    with tildes.
    Combining the result \eqref{eq:TTDeformationJ} with the GKP/W dictionary \eqref{eq:GKP/W} leads to an expression for the on-shell action of the gravitational system dual to the \TT-deformed CFT  
    \begin{align}
    \label{eq:deformedGKPW} 
        W_{[\mu]}[\g, J] = \left. S^{(ren)\star}[\tilde{\g}^{(0)}, \phi^{(0)}] - \mu \int\limits_{\rho \to 0} \td^2 x \sqrt{-\tilde \g_{(0)}} T_{ij}^\text{BY} T_{kl}^\text{BY} \gti_{(0)}^{i[k} \gti_{(0)}^{j]l} \right|_{\tilde{\gamma}_{ij}^{(0)} = \g_{ij}^{[0]}(\g_{ij}), \, \phi^{(0)} = J}
        \ .
    \end{align}
    As in \cref{eq:TTDeformationJ}, the function $\g_{ij}^{[0]}$ is given in terms of $\g_{ij}$ by solving \cref{eq:deformedMetricAndStressTensor},  
    where now $\braket{T_{ij}^{[0]}} \to T_{ij}^\text{BY}$ 
    is the asymptotic Brown--York stress tensor in this auxiliary geometry. We wrote out the $\cO_\TT$ operator explicitly with the appropriate metric contractions. 
    This expression is analogous to the holographic discussion of relevant and marginal double-trace deformations \cite{Klebanov:1999tb,Witten:2001ua,Berkooz:2002ug}, which change the boundary conditions of the appropriate bulk fields.

    The right-hand side of \cref{eq:deformedGKPW} is the solution to the variational problem with boundary conditions that fix $\g_{ij}$ rather than $\tilde{\gamma}_{ij}^{(0)}$, 
    \begin{equation}
    \label{MBC} 
	     \text{asymptotic mixed b.c. (MBC)}\; : \;\;\;\;\;\;\;	\delta \gamma_{ij}=0 .
    \end{equation} 
    This is guaranteed by \cref{eq:TmuDef} and \cref{eq:TTDeformationJ}, since \cref{eq:deformedGKPW} is simply the translation of the latter into bulk quantities.
    The boundary condition for the scalar is instead unchanged. 
     
    \par We end this section by giving explicit expressions for the auxiliary HMTZ solution. 
    Its line element is obtained from \cref{metriclinelement} or \cref{ESMbulkmetricFGgauge} by adding tildes to the boundary coordinates and the $B$ parameter\footnote{The precise relation between the auxiliary parameter $\tilde{B}$ and coordinates $(\tilde{t},\tilde{\varphi})$ and the original $B$ and $(t,\varphi)$ will be fully clarified in \cref{section:3.4}.} 
    \begin{align} \label{metriclinelementdef}
		\tilde{g}_{\alpha\beta}dx^{\alpha}dx^{\beta}&=-\left(\frac{\tilde{H}(r)}{\tilde{H}(r)+\tilde{B}}\right)^2\tilde{F}(r)d\tilde{t}^2+\left(\frac{\tilde{H}(r)+\tilde{B}}{\tilde{H}(r)+2\tilde{B}}\right)^2\frac{dr^2}{\tilde{F}(r)}+r^2d\tilde{\varphi}^2 
        \\
        &=\frac{d\rho^2}{4\rho^2}+\frac{1}{\rho}\left[\tilde{\gamma}^{(0)}_{ij}+O(\sqrt{\rho})\right]dx^idx^j=\frac{d\rho^2}{4\rho^2}+\frac{-d\tilde{t}^2+d\tilde{\varphi}^2}{\rho}+O(\rho^{-1/2}) \label{metriclinelementdefFG}
    \end{align}
    with 
    \begin{equation}
        \tilde{H}(r)=\frac{1}{2}\left(r+\sqrt{r^2+4\tilde{B}r}\right), \qquad \tilde{F}(r)=\big(\tilde{H}(r)\big)^2 - \nut \left(3\tilde{B}^2+\frac{2\tilde{B}^3}{\tilde{H}(r)}\right).
    \end{equation} 
    The asymptotic Brown--York stress tensor is 
    \begin{align} \label{T0}
        8\pi G \braket{T_{ij}^{[0]}} &= \tilde{\gamma}_{ij}^{(2)} - \left( c_4 + \frac{16}3 \right) \tilde{B}^2 \tilde{\gamma}_{ij}^{(0)} ,
    \end{align}
    which can be written as $\langle T_{ij}^{[0]}\rangle \td x^i \td x^j =\tilde{\mathcal{L}}_t d\tilde{t}^2 + \tilde{\mathcal{L}}_{\varphi}d\tilde{\varphi}^2$, where 
    \begin{align} \label{tildeL}
        \tilde{\cL}_t &\equiv \frac{\tilde{B}^2}{8\pi G} \left( c_4 - \frac23 + \frac{3\nut}2 \right)
         \ , &
        \tilde{\cL}_\varphi &\equiv -\frac{\tilde{B}^2}{8\pi G} \left( c_4 - \frac23 - \frac{3\nut}2 \right)
        \ .
    \end{align}
    Inverting \cref{eq:deformedMetricAndStressTensor} we have $\gamma_{ij}^{[\mu]}=\gamma_{ij}^{[0]}-2\mu \langle \hat{T}_{ij}^{[0]}\rangle+\mu^2 \langle \hat{T}^{[0]}_{ik}\rangle\langle\hat{T}^{[0]}_{jl}\rangle\gamma^{kl}_{[0]}$. Thus, we find 
    \begin{align}
    \label{eq:HMTZgammamu}
        \g_{ij}^{[\mu]} \td x^i \, \td x^j &= - ( 1 + \mu \tilde{\cL}_\varphi )^2 \td \tilde{t}^2 + ( 1 - \mu \tilde{\cL}_t )^2 \td \tilde{\varphi}^2
        \ .
    \end{align} 
    As we will see in \cref{section:3.4}, requiring that $\gamma^{[\mu]}_{ij}=\gamma_{ij}$ will fix the relation between the original coordinates $(t,\varphi)$ and the auxiliary coordinates $(\tilde{t},\tilde{\varphi})$.
    
    For $\nut = 0$, the metric in \cref{eq:HMTZgammamu} is simply proportional to the Minkowski one $\eta_{ij}$, whereas in the minimal subtraction scheme $c_4 = 2/3$ it reduces to
    \begin{align}
        c_4 &= \frac23:
        &
        \g_{ij}^{[\mu]} \td x^i \, \td x^j &= - (1 + \mu \tilde{L})^2 \td \tilde{t}^2 + (1 - \mu \tilde{L})^2 \td \tilde{\varphi}^2
        \ ,
    \end{align}
    where $\tilde{L} \equiv 3 \tilde{B}^2 \nut / 16\pi G$.
    Setting both $\nut = 0$ and $c_4 = 2/3$ leads to a trivial flow since $\g_{ij} = \g_{ij}^{[0]} = \eta_{ij}$.

    We can also express the hatted stress tensor of the deformed theory as
    \begin{align}
		\langle\hat{T}_{ij}^{[\mu]}\rangle \td x^i \, \td x^j &= \langle \hat{T}_{ij}^{[0]}-\mu \hat{T}^{[0]}_{ik}\hat{T}^{[0]}_{jl}\gamma^{kl}_{[0]} \rangle\td x^i \, \td x^j
		\nonumber \\
        &= \tilde\cL_\varphi (1 + \mu \tilde\cL_\varphi) \td \tilde t^2 + \tilde\cL_t (1 - \mu \tilde\cL_t) \td \tilde\varphi^2
        \ . 
    \end{align} 
    Given the relation $T_{ab}^{[\mu]}=\hat{T}_{ab}^{[\mu]}-\gamma_{ab}^{[\mu]}\hat{T}^{[\mu]}$ between the (deformed) hatted and non-hatted stress tensors, we can determine the following explicit form for the deformed stress tensor 
    \begin{align}
    \label{eq:HMTZstresstensor}
        \langle T_{ij}^{[\mu]} \rangle\td x^i \, \td x^j
        &= \frac{\tilde\cL_t (1 + \mu \tilde\cL_\varphi)^2}{1 - \mu \tilde\cL_t} \td \tilde{t}^2 + \frac{\tilde\cL_\varphi (1 - \mu \tilde\cL_t)^2}{1 + \mu \tilde\cL_\varphi} \td \tilde\varphi^2 \ .
    \end{align} 
    In the minimal subtraction scheme, the stress tensor again simplifies because $\tilde\cL_\varphi, \tilde\cL_t \to \tilde L$, whereas for $\nut = 0$ it is again proportional to $\eta_{ij}$.

    \subsection{Backreaction dispels the mirage} \label{section:3.1.1}  
    
    It was observed in \cite{Guica:2019nzm} that for the special case of pure gravity, the metric $\g^{[\mu]}_{ij}$ coincides with the rescaled induced metric on a surface of constant $\rho = \rho_0$ in the auxiliary 
    bulk geometry with asymptotic metric $\tilde\g_{ij}^{(0)} = \g_{ij}^{[0]}(\g)$, where $\rho_0 = -\mu / 4\pi G$. 
This was called the mirage of a bulk cutoff, referring to the bulk cutoff in the holographic $T\bar T$ proposal of \cite{McGough:2016lol}. 
The argument does not hold however when matter backreacts on the geometry, since the metric $\g_{ij}^{[\mu]}$ in \cref{eq:HMTZgammamu} is not proportional to the rescaled induced metric of the bulk \eqref{metriclinelementdef}.
    We can see this explicitly as follows: if we would require the $\tilde t \tilde t$ components to agree, $\tilde\g_{\tilde t \tilde t}^{[\mu]} = \rho _0 h_{\tilde t \tilde t}(\rho_0)$, we find the relation  
    \begin{align} \label{gammattmuxi} 
       \mu(\rho_0) = \frac1{\tilde\cL_\varphi} \left( -2\tilde B \sqrt{\rho_0} + \frac{\tilde B^2}4 (4-3\nut) \rho_0 + \ldots \right),
    \end{align} 
    whereas insisting that $\tilde{\g}_{\tilde \varphi \tilde \varphi}^{[\mu]} = \rho _0 h_{\tilde \varphi \tilde \varphi}(\rho_0)$, we find instead
    \begin{align} \label{gammaphiphimuxi} 
       \mu(\rho_0) = \frac1{\tilde\cL_t} \left( 2\tilde B \sqrt{\rho_0} - \frac{\tilde B^2}4 (4 + 3\nut) \rho_0 + \ldots \right).
    \end{align} 
    Therefore, we conclude that fixing the deformed metric does not correspond to a Dirichlet condition at finite radius for $\nut \neq 0$. 
    Curiously, for the special case $\nut = 0$ (and $c_4\neq \frac{2}{3}$, otherwise there is no flow), the perturbative relations \eqref{gammattmuxi}-\eqref{gammaphiphimuxi} reduce to the exact form
    \begin{equation} \label{muxinu-1}
        \nut = 0 \; \wedge \; c_4\neq \frac{2}{3}: \quad
        \gamma^{[\mu]}_{ij} dx^i dx^j = \rho_0 h_{ij}(\rho_0)dx^i dx^j \Leftrightarrow
        \mu(\rho_0) = \frac{8\pi G}{\tilde B^2 (c_4-\frac23)} (2 \tilde B \sqrt{\rho_0} - \tilde B^2 \rho_0). 
    \end{equation} 
    One may be tempted to interpret the above equation as the statement that, for a non-minimal choice of subtraction scheme, the \TT-deformed $\nut=0$ theory is dual to bulk gravity with a Dirichlet boundary condition at finite radius. However, in order to match the deformed one-point functions by equating the on-shell gravitational action with a Dirichlet boundary at finite cutoff to the \TT-deformed generating functional of the dual $\text{CFT}_2$, we find that one would need to introduce an infinite series of (asymptotically irrelevant) boundary terms, which require a non-universal fine-tuning depending on which one-point function one is interested to calculate. Therefore, there is no concrete sense in which we can establish a duality between the \TT-deformed $\nut = 0$ theory and the bulk gravity with a Dirichlet boundary condition at finite radius.

\subsection{Deformed energy} \label{section:3.4} 
    In this section we apply our holographic dictionary to verify that the energy spectrum for the \TT-deformed $\text{CFT}_2$ agrees with the well-known formula for the \TT-deformed energy \cite{Dubovsky:2012wk,Caselle:2013dra,Smirnov:2016lqw, Cavaglia:2016oda}.
    In doing so, it will be crucial to understand the precise map between the bulk solutions and the corresponding boundary states. 
    Note that the bulk HMTZ solutions have two free parameters, $\nu$ and $B$. The $\nu$ parameter enters in the action \eqref{action} through the matter potential, thus distinguishing different theories rather than different states. We instead identify the integration constant $B$ as the parameter labeling the state. We introduced $\tilde{B}$ in the auxiliary bulk solution 
    \eqref{metriclinelementdef} 
    to indicate that while 
    $B$ labels the state of the undeformed theory, $\Bt$ labels the bulk geometry corresponding to the same state in the deformed theory. 

    We begin by calculating the energy in the undeformed theory. 
    The line element of the conformal boundary of the HMTZ solution \eqref{ESMbulkmetricFGgauge}, which is identified via the standard AdS/CFT dictionary with the background geometry of the dual (undeformed) $\text{CFT}_2$, is
    \begin{equation} \label{ESMundeformedgeo}
		\gamma_{ij}dx^idx^j = \gamma_{ij}^{(0)}dx^idx^j = -dt^2+d\varphi^2,
    \end{equation} 
    describing an infinite Lorentzian cylinder with non-compact time coordinate $t\in(-\infty,\infty)$, compact spatial coordinate $\varphi\in[0,1)$ (we set the periodicity of the angular coordinate to unity, i.e. $\varphi \sim \varphi + 1$) and circumference $\mathcal{R}=1$.
    The undeformed energy $E_{[0]}$ of the dual $\text{CFT}_2$ is simply given by the integral over a fixed $t=t_0$ time-slice of the undeformed stress tensor \eqref{generalonepointfcn} contracted with the unit timelike normal vector 
    $u^{i}=\gamma^{ij}u_j=\delta^i_t$, yielding%
    \footnote{We should point out that this definition of the field-theory energy uses the unit normal with respect to $\g_{ij}$ and not with respect to the bulk induced metric $h_{ij}$. The distinction between the unit normal in the bulk and on the boundary is crucial when comparing to other notions of energy and mass such as in \cite{Balasubramanian:1999re,Blacker:2024rje}.}
    \begin{equation} \label{ESML} 
        E_{[0]} = \int_0^1 \td \varphi \sqrt{\g_{\varphi\varphi}^{(0)}} u^i u^j \langle T_{ij} \rangle = \cL_t
        \ .
    \end{equation} 
    In the minimal subtraction scheme, once the standard periodicity of the angular coordinate is restored, the undeformed energy indeed matches the black hole's mass \eqref{ESMMBH}, 
    \begin{equation} \label{ESMundeformedEminimal}
		c_4=\frac{2}{3}: \; \; \; \; \; \; \; \;  E_{[0]}=\frac{3B^2 \nut}{16\pi G}=\frac{M_{BH}}{2\pi}.
    \end{equation} 
    Since this state has a horizon, it carries an associated Bekenstein--Hawking entropy proportional to $r_h$ in \eqref{ESMundeformedrh}, 
	\begin{equation} \label{ESMundeformedS}
		\mathcal{S}_{[0]}=\frac{A^{[0]}_h}{4 G}=\frac{1}{4G}\int^{1}_0d\varphi\sqrt{g_{\varphi\varphi}}\Big|_{r=r_h}=\frac{ r_h}{4 G}=\frac{\Theta_{\nu}}{4G}B \ .
	\end{equation}

    To compare with the deformed field theory, we use the auxiliary bulk solution with metric $\tilde{g}_{\alpha\beta} dx^{\alpha} dx^{\beta}$ given in \cref{metriclinelementdef}. 
    Since we require the deformed field theory to live on the same background metric 
    $\g_{ij} \td x^i \, \td x^j = -\td t^2 + \td \varphi^2$, we know from \cref{sec:TTbulkDer} that this must equal the auxiliary metric 
    $\g_{ij}^{[\mu]}(\gti^{(0)}, \gti^{(2)})$, with 
    $\tilde{\g}^{(0)}_{ij}dx^idx^j=-d\tilde{t}^2+d\tilde{\varphi}^2$.
    (This is the reason we are introducing a separate auxiliary bulk metric.)
    Using \cref{eq:HMTZgammamu}, we find the following relation between the auxiliary bulk time and angular coordinates $(\tti, \varphit)$ and the field theory coordinates $(t, \varphi)$,   
    \begin{align}
    \label{ESMnewcoords}
        \tti &= \frac{t}{1 + \mu \cLt_\varphi}
        \ , &
        \varphit &= \frac{\varphi}{1 - \mu \cLt_t}
        \ .
    \end{align} 
    We read off that the periodicity of $\varphit$ is not 1 but $1 / (1 - \mu \cLt_t)$. 
  The deformed stress tensor \eqref{eq:HMTZstresstensor} on the auxiliary HMTZ solution is
    \begin{align} \label{ESMdeformedTLnewcoords} 
        \langle T^{[\mu]}_{ij} \rangle \td x^i \, \td x^j
        &= \frac{\cLt_t (1 + \mu \cLt_\varphi)^2}{1 - \mu \cLt_t} \td \tti^2 + \frac{\cLt_\varphi (1 - \mu \cLt_t)^2}{1 + \mu \cLt_\varphi} \td \varphit^2
        \nonumber \\
        &= \frac{\cLt_t}{1 - \mu \cLt_t} \td t^2 + \frac{\cLt_\varphi}{1 + \mu \cLt_\varphi} \td \varphi^2 \ .
    \end{align}
    The deformed energy $E_{[\mu]}$ is calculated as 
    \begin{align}
    \label{ESMEmu}
        E_{[\mu]} = \int_0^1 \td \varphi \sqrt{\g_{\varphi\varphi}} u^i u^j \langle T^{[\mu]}_{ij} \rangle = \frac{\cLt_t}{1 - \mu \cLt_t}
        \ .
    \end{align} 
    To compare this to the undeformed energy, we still need to find the relation between $\Bt$ and $B$ (or equivalently between $\cLt_t$ and $\cL_t$).
    We use the fact that the degeneracy of states does not flow under the \TT deformation and that this quantity is given by the exponential of the horizon area in Planck units.%
    \footnote{%
        An alternative way to derive the relation between $B$ and $\Bt$ is to introduce an auxiliary radial coordinate $\tilde{r}$ in the deformed geometry in \cref{metriclinelementdef}, and observe that the angular and radial components coincide exactly with the original one \eqref{metriclinelement} once we identify the coordinates as
        \begin{align}
            \varphi &= (1 - \mu \cLt_t) \varphit
            \ , &
            r &= \frac{\rt}{1 - \mu \cLt_t}
            \ ,
        \end{align}
        and indeed $B = \Bt / (1 - \mu \cLt_t)$. 
    } 
    The horizon of the black hole dual to the deformed theory is located at the Schwarz\-schild radial coordinate $\tilde{r}_h = \Bt \Theta_{\nu}$. Therefore, the deformed entropy $\mathcal{S}_{[\mu]}$ of the black hole dual to the deformed theory is given by 
    \begin{equation} \label{ESMdeformedS}
		\mathcal{S}_{[\mu]} = \frac{1}{4G}\int^{1 / (1 - \mu \cLt_t)}_0 d\varphit \sqrt{\tilde{g}_{\varphit\varphit}} \Big|_{r=\tilde r_h} 
        =\frac{\Theta_{\nu}}{4G}\frac{\Bt}{1-\mu \cLt_t},
    \end{equation} 
    where we used again the auxiliary HMTZ solution, given by \cref{metriclinelementdef}. 
    Comparing \cref{ESMundeformedS} and \cref{ESMdeformedS} we find
    \begin{align}
		B &=\frac{\Bt}{1-\mu \cLt_t} 
        \ , &
        &\text{i.e.}
        & 
		\cL_t &= \frac{\cLt_t}{\big(1-\mu \cLt_t\big)^2}.
    \end{align}
    Inverting the above relation and picking the branch that yields the correct undeformed limit, we have
    \begin{equation} \label{ESMLmuL}
		\cLt_t = \frac{1+2\mu \cL_t - \sqrt{1+4\mu \cL_t}}{2 \mu^2 \cL_t}.
    \end{equation}
    Therefore we find that, in terms of the undeformed energy \eqref{ESML}, the deformed energy \eqref{ESMEmu} takes the following form 
    \begin{equation}
		E_{[\mu]}=\frac{1}{2\mu}\left(\sqrt{1+4 \mu E_{[0]}}-1\right),
    \end{equation}
    which indeed is precisely the field-theoretic energy formula for a \TT-deformed $\text{CFT}_2$ with undeformed energy $E_{[0]}$ and vanishing angular momentum on the cylinder of unit circumference. 
    \par We observe that 
    calculating the mass of the black hole on the auxiliary bulk solution \eqref{metriclinelementdefFG} while instead imposing asymptotic DBC (i.e. as in \eqref{ESML}-\eqref{ESMundeformedEminimal}), one obtains 
    the same deformed energy 
    \begin{equation}
    c_4=\frac{2}{3}:  \;\; \; \; \; \; \; \; \; \; \; \frac{\tilde{M}_{BH}}{2\pi}=\int_0^{\frac{1}{1-\mu \tilde{\mathcal{L}}_t}}d\tilde{\varphi}\sqrt{\tilde{\gamma}^{(0)}_{\tilde{\varphi}\tilde{\varphi}}} \tilde{u}^i \tilde{u}^j \langle T^{[0]}_{ij}\rangle=E_{[\mu]},
    \end{equation}
    where we used $\tilde{u}^i=\delta^i_{\tilde{t}}$ and $\langle T_{ij}^{[0]}\rangle \td x^i \td x^j =\tilde{\mathcal{L}}_t d\tilde{t}^2 + \tilde{\mathcal{L}}_{\varphi}d\tilde{\varphi}^2$. 
This leads to a flow of the black hole's mass $M_{BH}=2\pi E_{[0]}\rightarrow \tilde{M}_{BH}=2\pi E_{[\mu]}$, which we can view as a holographic interpretation to the auxiliary flow (i.e. the green trajectory in \Cref{fig:flows}). 
    \par
\begin{figure}[ht] 
        \centering
        \includegraphics[draft=false, width=0.90\textwidth]{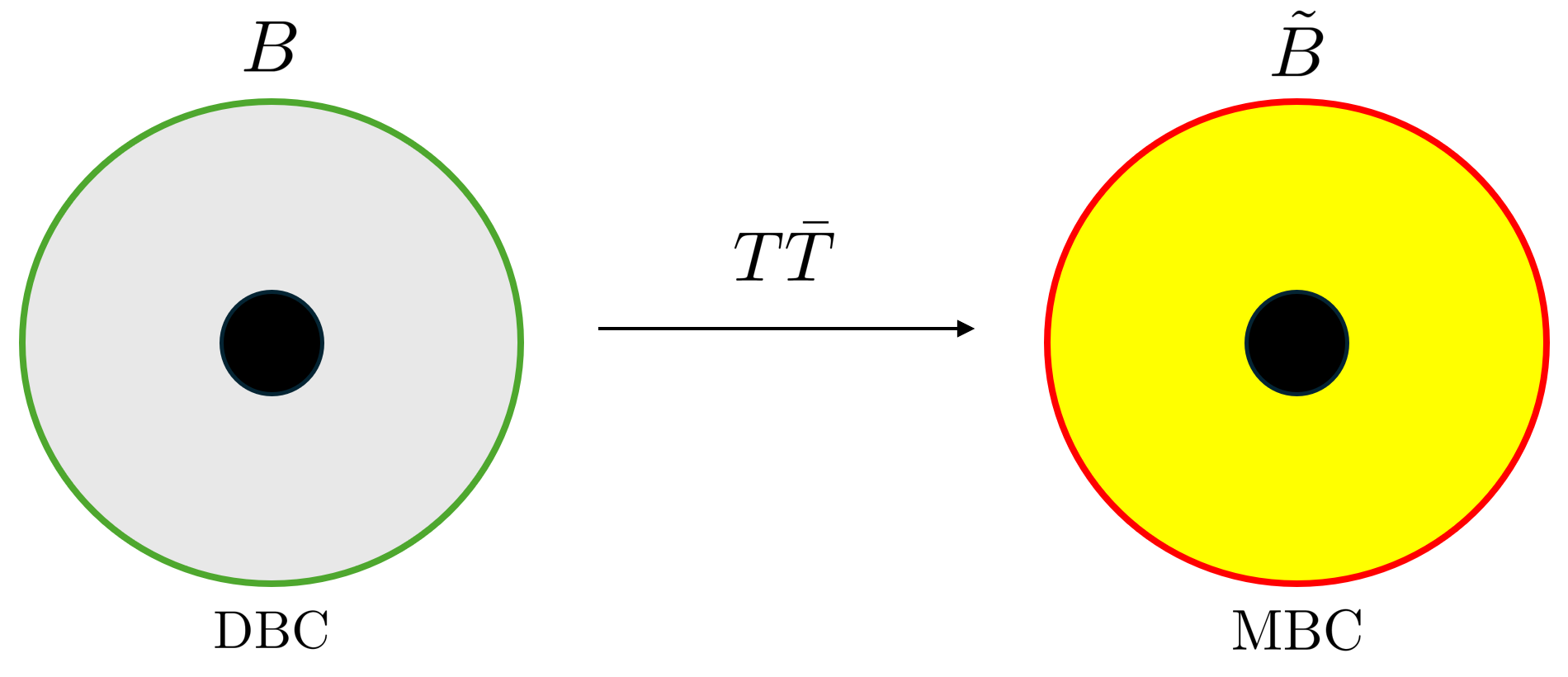}
        \caption{A schematic representation for the holographic interpretation of the \TT deformation in terms of the asymptotic mixed boundary conditions (MBC), applied in particular to the HMTZ solution. On the left, we have an equal-time slice of the HMTZ solution \eqref{metriclinelement} with asymptotic Dirichlet boundary conditions (DBC) on the metric, dual to the holographic $\text{CFT}_2$ discussed in \cref{subsection:2.1}. Deforming this $\text{CFT}_2$ with \TT is holographically dual to changing the asymptotic DBC to the asymptotic MBC \eqref{MBC}, as depicted on the right. The deformation can be interpreted in terms of the flow of the state parameter ($B\rightarrow \tilde{B}$) labeling the corresponding bulk solution.
        }
    \label{fig:DBCMBCFig}
    \end{figure}

 We include a graphical representation for the holographic interpretation of the $T\bar{T}$-deformation in terms of MBC in \Cref{fig:DBCMBCFig}.

\section{Dirichlet deformation}
\label{sec:Dirichlet}
In this section we derive the deformation of the holographic $\text{CFT}_2$ dual to the HMTZ model that results from imposing Dirichlet conditions at finite radial cutoff. As opposed to the previous section, we use the bulk to define the deformation and derive the corresponding field theory flow, similar to e.g.~\cite{Taylor:2018xcy,Shyam:2018sro,Hartman:2018tkw,Araujo-Regado:2022gvw}. 
We fix a particular bulk solution and define the deformed field theory generating functional as the on-shell gravitational action evaluated with a boundary at finite radial distance, identifying the background metric and scalar source of the deformed theory with the (rescaled) induced metric and scalar field on the cutoff surface.
We then translate the Hamiltonian equations governing the radial evolution of the bulk fields into a flow equation for the boundary theory. We identify the deforming operator as a specific combination of the \TT operator, the scalar operator, and their respective sources. This analysis reveals that the deformation is governed by the generator of Weyl transformations, consistent with \cite{Araujo-Regado:2022gvw} (in the semi-classical limit).
Finally, we compute the deformed energy via a bulk calculation and show that it satisfies a flow equation in the boundary theory that provides a natural generalization of the well-known \TT-deformed energy equation. 

\subsection{A bulk dual with Dirichlet boundary conditions at finite cutoff} \label{subsection: Dirichlet dictionary}

    The metric \eqref{ESMbulkmetricFGgauge} has an ADM form with vanishing shift vector
    \begin{equation} \label{metricdirichletadm}
		g_{\alpha\beta}dx^{\alpha}dx^{\beta} = 
        N^2(\rhot) d\rhot^2 + h_{ij}(\rhot, x^k) dx^i dx^j
        \ ,
    \end{equation}
    where we now allow for a general radial coordinate $\rhot$, demanding only that it agrees with the FG coordinate $\rho$ asymptotically: $N(\rhot) \to 1/ (2\rhot)$ as $\rhot \to 0$.
    
    The deformed boundary theory is defined by identifying its generating functional with the renormalized gravitational on-shell action $S_{[\rhot_0]}^{(ren)\star}$ \eqref{Srengeneral}, with $c_2=2$ and $c_4$ an arbitrary renormalization scheme-dependent constant
    \begin{align}
    \label{eq:WDirDef}
        W_{[\rhot_0]}[\g, J] &\equiv \left. S_{[\rhot_0]}^{(ren) \star}[h(\rhot_0), \phi(\rhot_0)] \right|_{h_{ij}(\tilde{\rho}_0) = \rhot_0^{-1} \g_{ij}, \;\phi (\tilde{\rho}_0)  = \rhot_0^{1/4} J}
        \ ,
    \end{align}
    and with Dirichlet boundary conditions at the finite radius $\rhot = \rhot_0$
    \begin{equation} \label{DBC}
		\text{Dirichlet b.c. (DBC):} \; \; \; \; \; \; \delta h_{ij}\big(\rhot_0\big)=0, \; \; \; \; \; \; \; \; \; \; \; \; \delta \phi \big(\rhot_0\big)=0.
	\end{equation} 
    In other words, the dual deformed boundary theory lives on the background metric $\gamma_{ij}\equiv \gamma_{ij}^{[\tilde{\rho_0}]}=\tilde{\rho}_0 h_{ij}(\tilde{\rho}_0)$ and is coupled to the scalar source $J\equiv J_{[\tilde{\rho}_0]}=\tilde{\rho}_0^{-1/4}\phi(\tilde{\rho}_0)$.
    Given the above dictionary, we can identify the deformed stress tensor with the renormalized Brown--York stress tensor $T_{ij}^\text{BY}$
    evaluated at finite radius%
    \footnote{%
        We raise indices with the inverse of the field theory metric $\gamma^{ij}$ and of the induced metric $h^{ij}\big(\rhot_0\big)$ for the former and the latter respectively, so we also have
        \begin{equation} \label{Tbdirichletup}
    		 T^{ ij}_{[\rhot_0]} = \rhot_0^{-2} T_\text{BY}^{ij} \Big|_{\rhot=\rhot_0}, \; \; \; \; \; \;  T_{[\rhot_0]} = \rhot_0^{-1} T_\text{BY}\Big|_{\rhot=\rhot_0},
        \end{equation}
        with $T_{[\rhot_0]} \equiv \gamma^{ij}T_{ij}^{[\rhot_0]}$ and $T_\text{BY}\equiv h^{ij} T_{ij}^\text{BY}$.
    }
    \begin{equation}
    \label{Tbdirichlet}
		\braket{ T_{ij}^{[\rhot_0]} } = -\frac{2}{\sqrt{-\gamma_{[\tilde{\rho}_0]}}} \frac{\delta W_{[\rhot_0]}}{\delta \gamma^{ij}_{[\tilde{\rho}_0]}}=-	\frac{2}{\sqrt{-h}}\frac{\delta S_{[\rhot_0]}^{(ren)\star}}{\delta h^{ij}}\Bigg|_{\rhot=\rhot_0}=\, T_{ij}^\text{BY} \Big|_{\rhot=\rhot_0}.
    \end{equation}
    Moreover, we can relate the scalar operator dual to the bulk matter field with the renormalized canonical momentum conjugate to $\phi$ evaluated at finite radius as follows
    \begin{equation} \label{Opidirichlet}
		\braket{ \mathcal{O}_{[\rhot_0]} } = \frac{1}{\sqrt{-\gamma_{[\tilde{\rho}_0]}}} \frac{\delta W_{[\rhot_0]}}{\delta J_{[\rhot_0]}}=\frac{\rhot_0^{-3/4}}{\sqrt{-h}}\frac{\delta S_{[\tilde{\rho}_0]}^{(ren)\star}}{\delta \phi}\Bigg|_{\tilde{\rho}=\tilde{\rho}_0} = \frac{\rhot_0^{-3/4}}{\sqrt{-h}} \pi_{\phi}^{(ren)}\Big|_{\tilde{\rho}=\tilde{\rho}_0} .
    \end{equation}
    On the HMTZ solutions, the renormalized Brown--York stress tensor and canonical momentum conjugate to the scalar are respectively given by
    \begin{align} \label{eq:TBY}
        T_{ij}^\text{BY} 
        = -\frac{1}{8\pi G}\Big[K_{ij}-\big(K-1-c_2\phi^2-c_4 \phi^4\big)h_{ij}\Big]
    \end{align}
    and
    \begin{equation}
    \label{eq:piRen}
		\pi_{\phi}^{(ren)}
        =-\frac{\sqrt{-h}}{\pi G}\left(\frac{\phi'}{N}+\frac{c_2}{4}\phi+\frac{c_4}{2}\phi^3\right)
    \end{equation}
    where $\phi' \equiv \pd_\rhot \phi$.
      
\subsection{Boundary flow equation and identification of the deforming operator} \label{subsection:floweqsourown}
    To compare with the \TT deformation, we want to identify the operator on the boundary that generates the flow defined by \cref{eq:WDirDef}.
    To make sure it is phrased in terms of the boundary fields and their canonical conjugates (after all, \TT is defined in terms of the conjugate to the boundary metric), we employ the radial Hamiltonian formalism.
    We first use the contracted Gauss--Codazzi equation $R=R[h]+K^2-K_{ij}K^{ij}+2\nabla_{\alpha}\left(n^{\beta}\nabla_{\beta}n^{\alpha}-n^{\alpha}K\right)$ to rewrite the gravitational action as 
    \begin{equation}
  	 S^{(ren)}_{[\rhot_0]}[g,\phi]=\int^{\rhot_0} d\rhot \; L^{(ren)}_{[\rhot]},
    \end{equation}
    with the renormalized radial Lagrangian given by
    \begin{align} \label{radiallagrangiandirichlet}
  	    L^{(ren)}_{[\rhot]}=\frac{N}{\pi    G}\int d^2x \sqrt{-h}\Bigg[&\frac{R[h]+K^2-K_{ij}K^{ij}}{16}-\frac{\left(\phi^{\prime}\right)^2}{2 N^2}-\frac{1}{2}h^{ij}\nabla_i\phi \nabla_j \phi-V_{\nu}(\phi) \nonumber
  	\\
  	     &-\frac{K}{8}\big(1+c_2\phi^2+c_4\phi^4\big)-\left(\frac{c_2}{4}\phi+\frac{c_4}{2}\phi^3\right)\frac{\phi'}{N}\Bigg].
    \end{align}
    Given the renormalized canonical momenta conjugate to the induced metric,
    \begin{align}
        \pi^{ij}_{(ren)} = \frac{\delta L^{(ren)}_{[\rhot]}}{\delta h_{ij}'} = -\frac{\sqrt{-h}}{16\pi G}\Big[K^{ij}-\big(K-1-c_2\phi^2-c_4 \phi^4\big)h^{ij}\Big] = \frac{\sqrt{-h}}2 T_\text{BY}^{ij}
    \end{align}
    and conjugate to the scalar field \eqref{eq:piRen},
    the radial Hamiltonian $H$ is then given by the Legendre transform
    \begin{equation} \label{radialHamdirichlet}
  	       H=\int d^2 x \left(\pi^{ij}_{(ren)}h_{ij}' + \pi^{(ren)}_{\phi}\phi'\right)-L^{(ren)}_{[\rhot]}=\int d^2 x \sqrt{-h} N\big(\rhot\big) \mathcal{H} \ ,
     \end{equation}
    with
    \begin{align}
    \label{eq:H}
        \mathcal H &= 
        \big(1+c_2\phi^2+c_4\phi^4\big)T_\text{BY}-4\pi G \left(	T_\text{BY}^{ij}T_{ij}^\text{BY}-T_\text{BY}^2 \right) - \frac{\pi G}{-2h} \left(\pi_{\phi}^{(ren)}\right)^2 \nonumber
  	    \\
  	    &\pheq  - \frac{1}{2 \sqrt{-h}}\left(\frac{c_2}{2}\phi+c_4\phi^3\right) \pi_{\phi}^{(ren)} + \frac1{\pi G} \left( -\frac{R[h]}{16} + \frac12 h^{ij}\nabla_i \phi \nabla_j \phi + \bar{V}_\nu(\phi) \right) \ .
    \end{align}
    Here we defined the modified potential
    \begin{align}
    \label{eq:tildeV}
        \bar{V}_\nu \equiv V_\nu + \frac18 \big(1+c_2\phi^2+c_4\phi^4\big)^2 - \frac18 \left(\frac{c_2}{2}\phi+c_4\phi^3\right)^2
        \ ,
    \end{align}
    which has a Taylor expansion that begins at order $\phi^6$ if we use the appropriate value for renormalization $c_2 = 2$, as the terms of order $\phi^0$, $\phi^2$ and $\phi^4$ cancel. 
    Going on-shell with respect to the bulk equations of motion enforces the Hamiltonian constraint
    $\mathcal{H}=0$. Using \cref{eq:H}, this can be rewritten as a flow equation for the trace of the Brown--York stress tensor, which we write in field theory terms using the dictionary \eqref{eq:WDirDef} ($\g_{ij}^{[\tilde{\rho}_0]} = \rhot_0 h_{ij}(\tilde{\rho}_0)$, $J_{[\tilde{\rho}_0]} = \rhot_0^{-1/4}\phi (\tilde{\rho}_0)$), as well as \cref{Tbdirichletup,Tbdirichlet,Opidirichlet} and the Brown--Henneaux central charge $c = 3 / 2G$.
    The result is 
    \begin{align} 
    \label{traceTTJ0gamma}
        \langle T_{[\tilde{\rho}_0]} \rangle
        = f\big(J_{[\tilde{\rho}_0]}\big) \bigg\langle \frac{6\pi \rhot_0}c \left( T^{ij} T_{ij} - T^2 \right) + \frac{3\pi}{4c} \sqrt{\rhot_0} \cO^2 + \frac12 \left(\frac{c_2}{2} + c_4 \sqrt{\rhot_0} J^2 \right) J \cO
        \nonumber \\
        + \frac{2c}{3 \pi} \Big( \frac{R[\g]}{16} - \frac12 \sqrt{\rhot_0} \g^{ij}\nabla_i J \nabla_j J - \frac1{\rhot_0} \bar{V}_\nu(\rhot_0^{1/4} J) \Big) \bigg\rangle_{[\tilde{\rho}_0]}  ,
    \end{align}
    where we defined $f(J) \equiv \big( 1 + c_2 \sqrt{\rhot_0}J ^2 + c_4 \rhot_0 J^4 \big)^{-1}$.
    As was remarked before \cite{McGough:2016lol}, the gravitational counterterm is essential in the derivation of this equation, as it provides the nonzero coefficient for the trace $T_{[\tilde{\rho}_0]}$.
    This trace equation furthermore contains all essential information about the deformation, as was stressed for the \TT deformation in \cite{Caputa:2020lpa} and for the Dirichlet deformation in \cite{Kruthoff:2020hsi}.
    Neglecting the matter counterterms, this equation would take a more standard form $\langle T_{[\tilde{\rho}_0]}\rangle = (6\pi \rhot_0 / c) \langle T^{ij} T_{ij} - T^2\rangle_{[\tilde{\rho}_0]} + \ldots$, reminiscent of the \TT deformation.
    However, we see that requiring a fully-renormalized undeformed limit, by including matter counterterms, alters this conclusion.

    When the deformation is turned off by taking the $\rhot \to 0$ limit, recalling \cref{eq:tildeV}, the trace flow equation \eqref{traceTTJ0gamma} for the renormalized theory reduces to
    \begin{equation}
      	c_2=2: \; \; \; \; \; \; \; \; \; \; \; \;\; \; \;  \langle T_{[0]}\rangle=\frac{c}{24\pi}R\left[\gamma_{[0]}\right]+\frac{1}{2}J_{[0]}\langle \mathcal{O}_{[0]}\rangle \ . \label{c22TaaRJO}
    \end{equation}
    Since $R[\g_{[\tilde{\rho}_0]}] = 0$ on the HMTZ solution, this reproduces the anomaly relation \eqref{anomalyrelationOT}.
    Using the trace flow equation \eqref{traceTTJ0gamma}, we can express the on-shell renormalized radial Lagrangian as follows
    \begin{align} \label{radiallagrangiandirichletonshell}
  	    L^{(ren)\star}_{[\tilde{\rho}]}=N\big(\tilde{\rho}\big)\int d^2x \sqrt{-h}\Big[&\big(1+c_2\phi^2+c_4\phi^4\big)T_\text{BY} \nonumber
        \\
        &-8\pi G \Big( T_\text{BY}^{ij}T_{ij}^\text{BY}-T_\text{BY}^2 \Big) \nonumber
  	\\
  	   &-\frac{\pi G}{-h} \left(\pi_{\phi}^{(ren)}\right)^2- \frac{1}{2 \sqrt{-h}}\left(\frac{c_2}{2}\phi+c_4\phi^3\right) \pi_{\phi}^{(ren)}\Big].
    \end{align}

    To derive a flow equation for the gravitational on-shell action and its holographic dual, the field theory generating functional, we must be careful about the quantities we keep fixed along the flow.
    As already remarked in \cite{Hartman:2018tkw}, it is natural from the field theory point of view to keep the background metric and scalar source fixed, even though these corresponds to the (rescaled) induced metric $h_{ij}(\rhot)$ and scalar field $\phi(\tilde{\rho}_0)$, which generically do not remain constant as we move the boundary inward.\footnote{%
    \label{fn:rhotSubtlety}%
        As we will see below, it is possible to avoid a non-trivial flow for the circumference of the background geometry by using the freedom of choosing the, so far undetermined, coordinate $\rhot$ as $1 / r^2$, where $r$ is the coordinate appearing in \cref{metriclinelement}.
    }
    We must therefore be careful to distinguish total derivatives from partial ones and to keep track of the quantities that are kept constant in the latter.
    The renormalized gravitational action evaluated on a 3d solution obeys the total derivative equation
    \begin{equation} \label{totaldvsgravequallonshell}
  	    \frac{d}{d\rhot_0}S^{(ren)\star}_{[\rhot_0]}\big[h(\rhot_0),\phi(\rhot_0) \big] = L^{(ren)\star}_{[\rhot_0]} .
    \end{equation}
    This can be split into explicit and implicit dependence on $\rhot_0$, as follows
    \begin{align}
        \frac{d}{d\rhot_0}S^{(ren)\star}_{[\rhot_0]} &= \partial_{\rhot_0}S^{(ren)\star}_{[\rhot_0]} + \int\limits_{\rhot=\rhot_0} d^2x \left(\frac{\delta S^{(ren)\star}_{[\rhot]}}{\delta h_{ij}} h_{ij}' + \frac{\delta S^{(ren)\star}_{[\rhot]}}{\delta \phi} \phi' \right) \nonumber
  	\\
  	     &= \partial_{\rhot_0}S^{(ren)\star}_{[\rhot_0]} + \int\limits_{\rhot=\rhot_0} d^2 x \left(\pi^{ij}_{(ren)} h_{ij}' + \pi^{(ren)}_{\phi} \phi' \right) ,
    \end{align}
    From this equation and \cref{radialHamdirichlet} we infer that the partial derivative is
    \begin{equation}
    \label{eq:partialRhoS}
        \partial_{\rhot_0}S^{(ren)\star}_{[\rhot_0]}\left[h(\rhot_0),\phi(\rhot_0)\right] = -H\big|_{\rhot = \rhot_0}.
    \end{equation}
    This partial derivative $\pd_{\rhot_0}$ is taken by keeping the other arguments of the function, in this case $h_{ij}$ and $\phi$, constant.
    In other words, this flow describes how the on-shell gravitational action changes as we move the boundary while insisting that $h_{ij}\big(\rhot_0\big)$ and $\phi\big(\rhot_0\big)$ are kept fixed. 
    This is not quite sufficient yet to give the flow of the generating functional $W_{[\rhot_0]}[\g, J]$, which is instead a function of the background metric $\gamma \equiv \gamma^{[\rhot_0]}_{ij}= \rhot_0 h_{ij}\big(\rhot_0\big)$ and scalar source $J \equiv J^{[\rhot_0]}= \rhot_0^{-1/4} \phi\big(\rhot_0\big)$.
    Its partial derivative $\pd_{\rhot_0}$ is taken by keeping $\g_{ij}$ and $J$ constant. 
    Therefore, using the holographic dictionary \eqref{eq:WDirDef} and the chain rule, we find
    \begin{align}
        \pd_{\rhot_0} W_{[\rhot_0]}[\g, J]
        &= \frac{d}{d\tilde{\rho}_0}W_{[\rhot_0]}- \int d^2 x \left( \frac{\d W_{[\tilde{\rho}_0]}}{\d \gamma^{[\tilde{\rho}_0]}_{ij}} \partial_{\rhot_0} \gamma_{ij}^{[\tilde{\rho}_0]} + \frac{\d W_{[\tilde{\rho}_0]}}{\d J_{[\tilde{\rho}_0]}}  \pd_{\rhot_0} J_{[\tilde{\rho}_0]} \right)
        \\
        &= L^{(ren)\star}_{[\tilde{\rho}_0]}- \int\limits_{\rhot=\rhot_0} d^2x \Bigg(\frac{\delta S^{(ren)\star}_{[\rhot]}}{\delta h_{ij}} h_{ij}' + \frac{\delta S^{(ren)\star}_{[\rhot]}}{\delta \phi} \phi' \Bigg) \nonumber
        \\
        &\pheq \qquad \qquad \qquad - \frac{1}{\tilde{\rho}_0}\int d^2 x \left( \frac{\d W_{[\tilde{\rho}_0]}}{\d \gamma^{[\tilde{\rho}_0]}_{ij}}  \gamma_{ij}^{[\tilde{\rho}_0]} - \frac{1}{4}\frac{\d W_{[\tilde{\rho}_0]}}{\d J_{[\tilde{\rho}_0]}}  J_{[\tilde{\rho}_0]} \right)
         \\
        &= -H \big|_{\tilde{\rho}=\tilde{\rho}_0}- \frac1{2 \rhot_0} \cW_{[\rhot_0]}\label{W-partial-flow}
    \end{align}
    where, in the intermediate step, we used that the second integral becomes proportional to the generator of Weyl transformations 
    \begin{equation} \label{Wweyl}
        \mathcal{W}_{[\rhot_0]}=\int d^2x \sqrt{-\gamma_{[\rhot_0]}}\left( \langle T_{[\rhot_0]}\rangle-\frac{1}{2}J_{[\rhot_0]}\langle\mathcal{O}_{[\rhot_0]}\rangle\right)
        \ .
    \end{equation}
    Since $H=0$ on-shell due to the radial Hamiltonian constraint, we conclude that the generating functional flows as 
    \begin{equation}
     \rhot_0 \pd_{\rhot_0} W_{[\tilde{\rho}_0]} [\gamma,J]= -\frac{\cW_{[\tilde{\rho}_0]}}{2} \ .
    \end{equation}
    In the asymptotic limit, we can use \eqref{c22TaaRJO} to recover the conformal anomaly of the renormalized dual $\text{CFT}_2$ 
    \begin{align}
       \lim\limits_{\rhot_0\rightarrow 0} \left(\rhot_0 \pd_{\rhot_0} W_{[\rhot_0]} \right)
       &=-\frac{c}{48\pi}\int d^2 x \sqrt{-\gamma_{[0]}} R[\gamma_{[0]}].
    \end{align}
    Combining this result with the trace flow equation \eqref{traceTTJ0gamma} \cite{Hartman:2018tkw}, we obtain the flow equation for the generating functional  of the boundary theory dual to the HMTZ model with Dirichlet condition at finite cutoff in terms of the deforming operator, 
    \begin{align} \label{partialrhoflow}
        &\rhot_0 \pd_{\rhot_0} W_{[\rhot_0]}[\g, J]
        \\
        &= -\frac12 \int d^2 x \sqrt{-\g_{[\tilde{\rho}_0]}} f\big(J_{[\tilde{\rho}_0]}\big) \, \Bigg\langle 
        \begin{aligned}[t]
            &\frac{6\pi \rhot_0}c \cO_\TT+ \frac{3\pi}{4c} \sqrt{\rhot_0} \cO^2 + \frac12 \left(\frac{c_2}{2} + c_4 \sqrt{\rhot_0} J^2 - \frac{1}{f(J)} \right) J \cO
            \\
            & + \frac{2c}{3 \pi} \Bigg( \frac{R[\g]}{16} - \frac12 \sqrt{\rhot_0} \g^{ij}\nabla_i J \nabla_j J - \frac1{\rhot_0} \bar{V}_\nu\big(\rhot_0^{1/4} J\big) \Bigg) \Bigg\rangle_{[\tilde{\rho}_0]}
            \ .
        \end{aligned}
        \nonumber
    \end{align}
    To compare to the previous section, we can define the rescaled deformation parameter $\mut = -4\pi G \rhot_0 = -6\pi \rhot_0 / c$ and use $c_2 = 2$,
    \begin{align}
    \label{eq:DirichletDeformation}
        & \pd_\mut W_{[\mut]}[\g, J]
        \\
        &= \frac12 \int d^2 x \sqrt{-\g} \, f(J) \Bigg\langle 
        \begin{aligned}[t]
            & \cO_\TT + \sqrt{-\frac{3\pi}{32 c \mut}} \cO^2 - \frac{1}{2\mut} \left( (c_4 - 2) \sqrt{-\frac{c \mut}{6\pi}} J^2 + \frac{c_4 c \mut}{6\pi} J^4 \right) J \cO
            \\
            & - \frac{2c}{3 \pi \mut} \left( \frac{R[\g]}{16} - \frac12 \sqrt{-\frac{c \mut}{6\pi}} \g^{ij}\nabla_i J \nabla_j J + \frac{6\pi}{c \mut} \bar{V}_\nu \right) \Bigg\rangle_{[\tilde{\mu}]}
            \ , 
        \end{aligned}
        \nonumber
    \end{align}
    where we renamed $\gamma_{ab}\equiv \gamma_{ab}^{[\tilde{\rho}_0]}\equiv \gamma^{[\tilde{\mu}]}_{ab}$, $J\equiv J_{[\tilde{\rho}_0]}\equiv J_{[\tilde{\mu}]}$, $\mathcal{O}_{[\tilde{\rho}_0]}\equiv \mathcal{O}_{[\tilde{\mu}]}$ and $T_{ab}^{[\tilde{\rho}_0]}\equiv T_{ab}^{[\tilde{\mu}]}$.
    As expected \cite{McGough:2016lol}, we see the \TT operator appear in this equation, but it is corrected by the addition of double-trace terms for the scalar operator and a polynomial in its source, as well as by the overall source-dependent prefactor which multiplies even the \TT part of the deforming operator.
    In the presence of matter, we therefore find a very explicit distinction between the \TT deformation \eqref{eq:TTpartial} and the Dirichlet deformation \eqref{eq:DirichletDeformation}.

    The equation \eqref{partialrhoflow} describes how $W$ changes with $\rhot_0$ as we keep $\g_{ij}$ and $J$ constant.
    For later reference, we note that keeping only the background geometry $\g_{ij}$ constant while allowing $J = \phi(\rhot_0) / \rhot_0^{1/4}$ to change as it does in the bulk, we find that $W$ changes as
    \begin{align} \label{dirichletflowJfree}
    \pd_{\rhot_0} W_{[\rhot_0]}[\g, J]
        + \int d^2 x  \left( \frac{\d W_{[\tilde{\rho}_0]}}{\d J_{[\tilde{\rho}_0]}}  \pd_{\rhot_0} J_{[\tilde{\rho}_0]}\right)=-\frac{1}{2\tilde{\rho}_0}\int d^2x \sqrt{-\gamma_{[\tilde{\rho}_0]}}\left(\langle T_{[\tilde{\rho}_0]} \rangle - P_{[\tilde{\rho}_0]}\right)
    \end{align}
    where we define
    \begin{equation}\label{PN}
     P = -N \rhot_0 \Braket{ \frac{3\pi}c \sqrt{\rhot_0} \mathcal{O}^2 +  \left( \frac{c_2}{2} + c_4 \sqrt{\tilde{\rho}_0} J^2 \right) J \mathcal{O} } \ ,
    \end{equation}
    which, as we will see below, can be interpreted as a matter contribution to the pressure.

\subsection{Deformed energy}

    We are now interested in calculating the deformed energy from both the bulk and boundary perspectives.
    Before doing so, we first address the subtlety mentioned in \cref{fn:rhotSubtlety}, namely the fact that the rescaled induced metric $\rhot_0 \, h_{ij}(\tilde{\rho}_0)$ does not remain constant in the bulk as we move the boundary. As we will see, this observation combines nicely with an ambiguity we have so far left unspecified: the freedom that remains to choose the radial 
    coordinate $\rhot$. 
    
    In a similar spirit to that of section \ref{section:3.4}, we regard \eqref{metriclinelement} as the undeformed bulk geometry, while we introduce an auxiliary bulk geometry of the form \eqref{metriclinelementdef}, with decorated boundary coordinates $(\tilde{t},\tilde{\varphi})$ and state parameter $\tilde{B}$.
    Hence, we read off  from \eqref{metriclinelementdef} that the rescaled induced metric on a $r=r_0$ surface within the auxiliary bulk geometry is
    \begin{align}
    \label{eq:gijDirichlet}
        \g_{ij}^{[\rhot_0]} dx^i \, d x^j
        &= \rhot_0 \left( -\frac{\tilde H^2 \tilde F}{(\tilde H  + \tilde B)^2} d\tti^2 + r_0^2 d\varphit^2 \right)\Bigg|_{r=r_0}
        \ .
    \end{align} 
    The gauge-invariant information contained in this rescaled induced metric consists of its Ricci scalar, which vanishes everywhere, as well as the circumference of the spatial circle. 
    Since we are to compare this with a field theory metric $\gamma_{ij}dx^{i}dx^{j}=-dt^2 + d\varphi^2$ with unit circumference ($\varphi\sim\varphi +1$), we see that for the specific choice
    \begin{align} \label{tilderhor}
        \rhot_0 &= 1 / r_0^2
        \ ,
    \end{align} 
    the radius of the spatial circle within the rescaled geometry remains unchanged along the flow into the bulk.
    This is furthermore a good choice of radial coordinate 
    because it agrees asymptotically with $\rho$ given in \cref{rrhogeneral}, so we will restrict to this case from here on.
    
\begin{figure}[ht] 
        \centering
        \includegraphics[draft=false, width=0.90\textwidth]{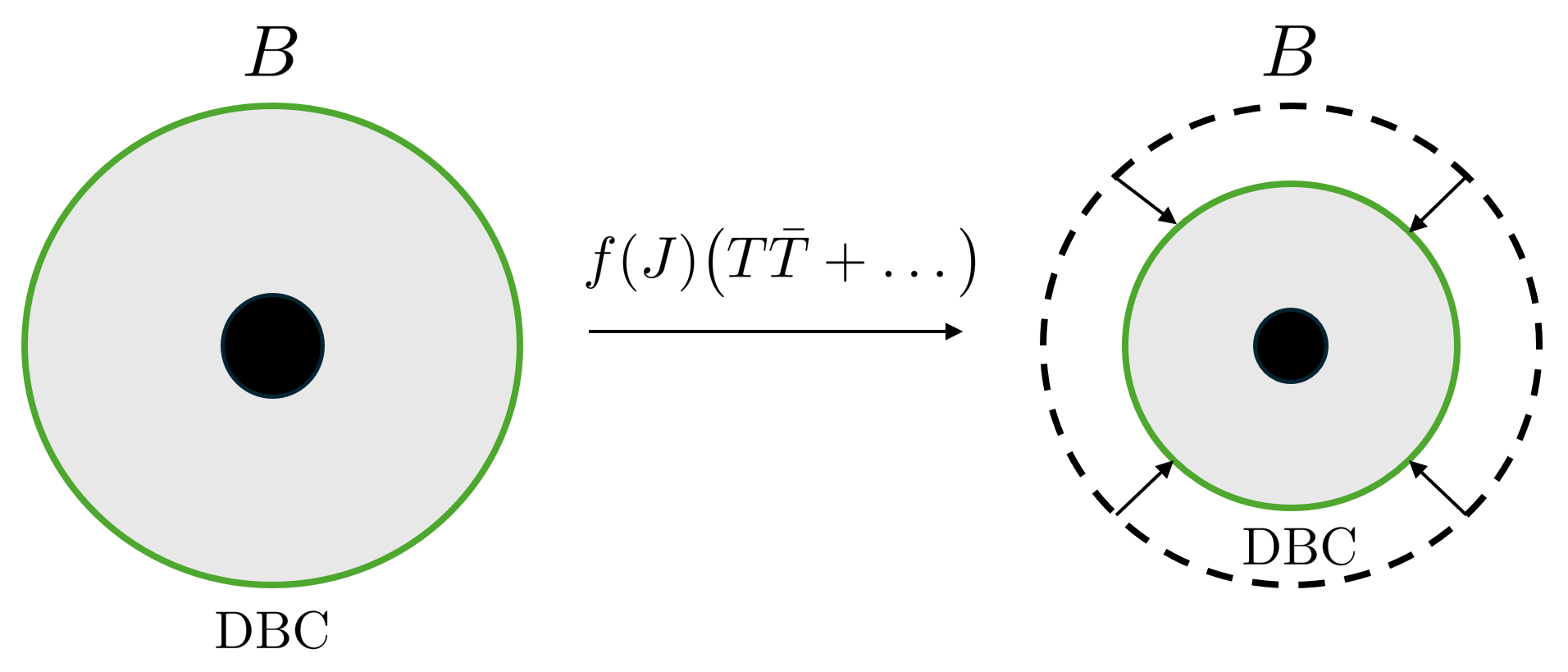}
        \caption{A schematic representation of the holographic interpretation of the Dirichlet deformation, applied to the HMTZ solution. On the left, we have the HMTZ solution \eqref{metriclinelement} with asymptotic DBC on the metric, dual to the holographic $\text{CFT}_2$ discussed in \cref{subsection:2.1}. Deforming this $\text{CFT}_2$ with $f(J)(\TT + \dots)$, as prescribed by the flow \eqref{partialrhoflow}, is holographically dual to changing the asymptotic DBC to a DBC at finite cutoff \eqref{DBC}, as depicted on the right. Due to our convenient choice of deformation coupling \eqref{tilderhor}, the state parameter of the HMTZ solution 
        is unchanged. 
        } 
    \label{fig:DBCDBCFig}
    \end{figure}

    Comparing \cref{eq:gijDirichlet} with the CFT metric, 
     we can then identify%
    \footnote{For more general bulk solutions with non-zero angular momentum,  
    $\tti$ will be still be proportional to $t$ but $\varphit$ will be identified with a linear combination of $t$ and $\varphi$, with coefficients determined by the absence of $g_{t\varphi}$.}
    \begin{align}
    \label{Direwcoords}
        \tti &= \frac{r_0\big(\tilde{H}+\tilde{B}\big)}{\tilde{H}\sqrt{\tilde{F}}}\Bigg|_{r=r_0}\,t
        \ , &
        \varphit &= \varphi.
    \end{align} 
    Analogously to section \ref{section:3.4}, we see that, as the deformation is turned off 
    $\rhot_0 \to 0$ and the cutoff surface within the auxiliary geometry is pushed to the asymptotic boundary, the decorated and undecorated coordinates coincide. Due to our convenient choice of radial coordinate $\tilde{\rho}$, the decorated and undecorated angular coordinates $\tilde\varphi$ and $\varphi$ actually coincide for any $\tilde{\rho}_0$ and so do their periodicities. 
    Furthermore, since the horizon does not change under the Dirichlet deformation, we can again equate the black hole entropies on the undeformed and auxiliary bulk geometries 
    \begin{equation}
      \mathcal{S}_{[\tilde{\rho}_0]}=\frac{1}{4G}\int_0^1d\tilde\varphi\sqrt{\tilde{g}_{\tilde{\varphi}\tilde\varphi}}\Big|_{r=\tilde{r}_h}=\frac{\tilde{r}_h}{4 G}=\frac{r_h}{4G}=\frac{1}{4G}\int_0^1d\varphi\sqrt{g_{\varphi\varphi}}\Big|_{r=r_h}=\mathcal{S}_{[0]}.
    \end{equation}
    Given the radial locations of the black hole horizons in the auxiliary and undeformed bulk geometries, $\tilde r_h=\tilde B \Theta_{\nu}$ and $r_h= B \Theta_{\nu}$ 
    respectively, we conclude that the state parameters match
    \begin{equation}
    B=\tilde{B}.
    \end{equation}
    The effect of the Dirichlet deformation on the bulk geometry is depicted in \Cref{fig:DBCDBCFig}.

\subsubsection{Gravitational derivation of the deformed energy and pressure} \label{subsection:gravitationaldeformedenergy}
    We now proceed to calculate how the energy, as measured in the field theory with metric $\g_{ij}$, changes under the deformation.
    It is given by%
    \footnote{We again use that the HMTZ solution has vanishing angular momentum. Otherwise the relation between $T_{tt}$ and $(T_{\tti \tti}, T_{\tti \varphit}, T_{\varphit \varphit})$ would be more complicated.}
    \begin{equation} \label{EmudefinitionDirichletEuclidean}
      	E_{[\rhot_0]} 
        = \int^1_0 d\varphi \sqrt{\gamma_{\varphi\varphi}}u^{i}u^{j} \langle T^{[\rhot_0]}_{ij} \rangle
        =\langle T^{[\rhot_0]}_{tt} \rangle
        = \frac{(H + B)^2}{\rhot_0 H^2 F} T^\text{BY}_{\tti \tti}\big|_{\tilde{\rho}=\tilde{\rho}_0}
        = -\frac{h_{\tti \tti}}{\rhot_0} T_\text{BY}^{\tti \tti}\big|_{\tilde{\rho}=\tilde{\rho}_0}
    \end{equation}
    where $u^{i} = \d_t^i$ is again the timelike unit normal to constant $t=t_0$ slices of the field theory's background geometry $\gamma_{ij}$, and $T^\text{BY}_{ij}$ is the Brown--York stress tensor \eqref{eq:TBY} on the auxiliary bulk geometry. We obtain the following expression for the deformed energy%
    \footnote{%
        In general, for a metric of the form $g_{tt}(r)dt^2+g_{rr}(r)dr^2+r^2d\varphi^2$, the deformed energy evaluates to $E_{[\tilde{\rho}_0]}=\frac{r_0^2}{8\pi G}\left(1-(r_0^2 g_{rr}(r_0))^{-1/2}\right)+\dots$, where the dots denote the possible matter counterterm contributions. For the non-rotating BTZ for example, we have $g_{rr}(r)=\frac{1}{r^2-8GM}$, hence the corresponding deformed energy is $E_{[\tilde{\rho}_0]}=\frac{r_0^2}{8\pi G}\left(1-\sqrt{1-\frac{8GM}{r_0^2}}\right)$, i.e. the \TT deformed energy on the cylinder of unit circumference.} 
    \begin{align}
     \label{deformedenergygrav}
      	E_{[\tilde{\rho}_0]}
        &=\frac1{8\pi G \rhot_0} \left( 1 - \sqrt{\rhot_0 F} \, \frac{H + 2B}{H + B} +c_2 \phi^2 + c_4 \phi^4 \right) \Bigg|_{\tilde{\rho}=\tilde{\rho}_0}
        \nonumber \\
        &= \frac{c}{12\pi\tilde{\rho}_0} \left( 1 - \sqrt{\rhot_0 F} \, \frac{H + 2B}{H + B} + c_2 \sqrt{\rhot_0} J^2 + c_4 \rhot_0 J^4 \right) \ ,
    \end{align}
    where the second line is just the same result in field theory language. In the case $\nut = 0$, this takes the simple form
    \begin{equation}
      	\nut = 0: \; \;\; \; \;\;\;\;\;\;\;\;\;\; E_{[\tilde{\rho}_0]} = \frac{c}{12\pi\tilde{\rho}_0} \left(1-\sqrt{1 +4B \sqrt{\rhot_0}} + c_2 \sqrt{\rhot_0} J^2 + c_4 \rhot_0 J^4  \right).
    \end{equation}
    As we turn off the deformation, pushing the cutoff surface back to the asymptotic boundary, the renormalized deformed energy can be expanded as
    \begin{align}
      	c_2=2: \; \; \; \; \; \; \; \lim\limits_{\tilde{\rho}_0\rightarrow 0}E_{[\tilde{\rho}_0]}
        &= \frac{c}{12\pi} \left[ B^2 \left( c_4 - \frac23 + \frac{3\nut}2 \right) + \sqrt{\tilde{\rho}_0} B^3 \left( \nut - \frac{8 c_4}3 + \frac{76}{45} \right) + \ldots \right]
    \end{align}
    which reproduces the mass of the black hole, i.e. $E_{[0]}=\frac{M_{BH}}{2\pi}$ with \eqref{ESMMBH}.

Before ending this section, we present a gravitational calculation of the pressure term $\braket{T_{\varphi \varphi}^{[\tilde{\rho}_0]}}$. 
This will justify the definition of $P$ in \cref{PN} as the matter contribution to the pressure.
Indeed, we find
\begin{equation}
\langle T_{\varphi \varphi}^{[\tilde{\rho}_0]}\rangle=T^{BY}_{\varphi\varphi}\big|_{\tilde{\rho}=\tilde{\rho}_0}=-\frac{1}{8\pi G \tilde{\rho}_0}\left(1+\frac{\tilde{\rho}^{3/2}\partial_{\tilde{\rho}}g_{tt}}{g_{tt}\sqrt{g_{rr}}}+c_2 \phi^2+c_4\phi^4\right)\Bigg|_{\tilde{\rho}=\tilde{\rho}_0}
,
\end{equation}
which, in terms of the deformed energy, takes the following form\footnote{It is useful to split $E_{[\tilde{\rho}_0]}=\bar{E}+E_{ct}$, with $E_{ct}$ the contribution from the matter counterterms ($c_2$ and $c_4$), and use $\bar{E}_{\tilde{\rho}_0}+2\tilde{\rho}_0\partial_{\tilde{\rho}_0}\bar{E}_{\tilde{\rho}_0}=-\frac{1}{8\pi G\tilde{\rho_0}}\left(1-\left(\tilde{\rho}_0/g_{rr}\right)^{3/2}\partial_{\tilde{\rho}}g_{rr}\right)\big|_{\tilde{\rho}=\tilde{\rho}_0}$ along with the useful HMTZ identities $g_{tt}=-\frac{1}{g_{rr}}\left(1+\frac{4B}{r}\right)^{-1}=-\frac{4 r^3 }{B}\frac{(\partial_r \phi)^2}{g_{rr}}$.}
\begin{equation} \label{TphiphiP}
  \langle T_{\varphi \varphi}^{[\tilde{\rho}_0]}\rangle=E_{[\tilde{\rho}_0]}+2\tilde{\rho}_0\partial_{\tilde{\rho}_0}E_{[\tilde{\rho}_0]}+P_{[\tilde{\rho}_0]},
\end{equation}
where $P$ is given by \cref{PN}. We can rewrite it also as
\begin{equation}\label{P}
P = \frac12 \left( \frac1{f(J)} - \frac{12 \pi \rhot_0 E}c \right)^{-1} \Braket{ \frac{3\pi}c \sqrt{\rhot_0} \mathcal{O}^2 +  \left( \frac{c_2}{2} + c_4 \sqrt{\tilde{\rho}_0} J^2 \right) J \mathcal{O} } \ .
\end{equation}
    
\subsubsection{Boundary flow equation for the energy}
\label{sec:dirEnergyBdyFlow}
    To derive by field theory arguments how the energy levels flow, we combine \cref{EmudefinitionDirichletEuclidean,TphiphiP} and find
    \begin{align}
        \langle T^{[\rhot_0]} \rangle= -\langle T_{tt}^{[\rhot_0]} \rangle + \langle T_{\varphi \varphi}^{[\rhot_0]}\rangle = 2\rhot_0 \pd_{\rhot_0} E_{[\tilde{\rho}_0]} + P_{[\rhot_0]}
        \ ,
    \end{align}
    or in other words,
    \begin{equation}
    \partial_{\tilde{\rho}_0}E_{[\tilde{\rho}_0]}=\frac{\langle T_{[\tilde{\rho}_0]}\rangle}{2\tilde{\rho}_0}-\frac{P_{[\tilde{\rho}_0]}}{2\tilde{\rho}_0}.
    \end{equation}
    Thus we conclude that the deformed energy flows like the deformed generating functional, with the background metric, but not the scalar source, kept constant, as in \cref{dirichletflowJfree}. 
    This generalizes the argument that the flow of the energy levels is just minus that of the action \cite{Kruthoff:2020hsi} to deformations for which sources $J$ do not remain fixed.
    
    Using the trace flow equation \eqref{traceTTJ0gamma} (with $R[\gamma]=\nabla_i J=0$ on the HMTZ solution), \cref{TphiphiP} and the explicit form of $P_{[\tilde{\rho}_0]}$ in \cref{P}, we arrive at the following differential equation for the deformed energy
     \begin{align}
     \label{eq:drhoE}
        \pd_{\rhot_0} E_{[\rhot_0]} 
        &= \frac{6\pi}c f\big(J_{[\rhot_0]}\big)  \left( E_{[\rhot_0]}^2 + 2 \rhot_0 E_{[\rhot_0]} \pd_{\rhot_0} E_{[\rhot_0]} - \Braket{\frac1{16 \sqrt{\rhot_0}} \cO^2 + \frac{c^2}{18\pi^2 \rhot_0^2} \bar{V}(\rhot_0^{1/4} J)}_{[\tilde{\rho}_0]} \right).
    \end{align}
    It is straightforward\footnote{Once again, it is easier to split the deformed energy as $E_{[\tilde{\rho}_0]}=\bar{E}+E_{ct}$, with $E_{ct}$ the contribution from the matter counterterms, and  first consider the unrenormalized case ($c_2=c_4=0$). Once the unrenormalized limit is solved, the addition of the contributions from $c_2$ and $c_4$ follows entirely algebraically. } to check that the energy \eqref{deformedenergygrav} provides a solution to \cref{eq:drhoE}. As a cross-check, it is easy to see that these equations reduce to their \TT analogues in the absence of matter for $\mut = -6
    \pi \rhot_0 / c$; in particular \cref{eq:drhoE} turns into the well-known Burgers' equation.

\section{Discussion}
\label{sec:discussion}
    We have worked out in detail how the HMTZ solution to the gravitational action \eqref{action} changes as we deform the system by imposing either mixed boundary conditions asymptotically, as in \cref{MBC}, or Dirichlet boundary conditions at finite radius \eqref{DBC}.
    Holographically, these deformations are dual to the \TT deformation \eqref{eq:TTpartial} and what we dubbed the “Dirichlet deformation” \eqref{partialrhoflow} of the generating functional $W$ of the field theory%
    \footnote{As mentioned in the introduction, we are really talking about the field theory dual to a UV completion for which \cref{action} is a low-energy approximation.}.
    As a cross check on our result, we have provided a bulk and a boundary calculation of how the energy spectrum changes for each of these deformations.

    Our result makes the distinction between these two deformations very clear, as we will summarize in \cref{sec:TTvsDirichlet}. It makes use of a concrete model for testing existing proposals in the literature both on (generalized) \TT deformations and on finite-volume Dirichlet boundary conditions in the presence of bulk matter,  without relying on perturbation theory and avoiding abstraction in favor of concreteness.  In \cref{sec:comparison} we discuss the comparison to known results in the literature. 
    We end with an outlook in \cref{sec:outlook}.

\subsection{\TT vs Dirichlet}
\label{sec:TTvsDirichlet}
    As was amply emphasized in the previous sections, the presence of matter in the bulk breaks the agreement between the \TT and the Dirichlet deformations. 
    The most striking difference, from the field theory point of view, is that the former may be regarded as a UV-complete (albeit non-Wilsonian) field theory whereas the latter should be treated as an effective field theory due to the presence of non-solvable irrelevant terms.
    
    Nevertheless, at the level we are working on --- classically in the bulk --- these deformations have some aspects in common.
    One of these is the fact that the energy levels flow exactly the same way (up to a sign) as the generating functional $W$.
    A similar statement was derived for the classical Hamiltonian using field theory considerations in \cite{Kruthoff:2020hsi}.
    Our result provides further support for this statement at the large-$c$ semiclassical level and even for deformations which consist of adding an operator and turning on a source at the same time.

    This leads us to a second similarity: comparing the bulk to the boundary perspective of either deformation requires quite some care with keeping the appropriate quantities constant.
    This is perhaps most striking for the flow of the gravitational on-shell action with Dirichlet boundary conditions if the induced metric and scalar field on the boundary are kept constant, as seen in \cref{eq:partialRhoS}: diffeomeorphism invariance in the bulk (reflected in the Hamiltonian constraint) imposes that the on-shell action does not change along this direction.
    Indeed, as was already remarked in \cite{Hartman:2018tkw}, the non-trivial part of the Dirichlet flow comes entirely from the fact that we keep the \emph{rescaled} induced metric $\g_{ij}$ and the source $J$, i.e. a rescaled version of the scalar field on the boundary, fixed in our prescription.
    A similar remark applies to the \TT deformation, for example when calculating the deformed energies using the gravitational prescription: one has to properly identify the auxiliary bulk geometry, including the correct relation between the bulk and boundary metric as well as the parameters ($B$ and $\Bt$) labeling the bulk solution corresponding to a given boundary state, in order to find the correct result. 
    
    Finally, the energy matching in \cref{sec:dirEnergyBdyFlow} made it clear that the field theory deformation only agrees with the bulk if one imposes that the source $J$ flows exactly in the way prescribed by the bulk scalar field profile. Otherwise, rather than describing parts of the same universe of decreasing size, the deformation traces out a nontrivial path in the bulk configuration space of solutions that have different scalar profiles.

\subsection{Comparison with the literature}
\label{sec:comparison}
    As already alluded to in the previous sections, there are many points of overlap and some differences between our Dirichlet result and the existing literature.
    For example, our choice to fix the \emph{rescaled} induced metric $\g_{ij}$ rather than the unrescaled one $h_{ij}$ differs from e.g. \cite{Kraus:2018xrn,Taylor:2018xcy, Shyam:2018sro}. However, we understand this as a matter of convention rather than an actual physical distinction: it should be possible to reproduce the same physics with either convention.
    
    Our choice of Dirichlet dictionary in \cref{subsection: Dirichlet dictionary} is the analogue of \cite{Hartman:2018tkw}. However, once the need for matter counterterms is taken into account, the resulting analysis yields modified flow equations for the trace of the deformed stress tensor and for the deformed generating functional. This leads to the additional scalar field contribution in the Weyl generator \eqref{Wweyl} which correctly reproduces the asymptotic Weyl anomaly of the undeformed CFT.
    
    Our analysis in \cref{subsection:floweqsourown} is consistent in the semi-classical limit with that of \cite{Araujo-Regado:2022gvw}, where the Dirichlet deformation in the presence of general bulk matter is discussed. Their renormalising counterterms and trace flow equation agree with ours up to the scheme-dependent term, for which they only consider the case $c_4 = 0$. The authors derive the irrelevant operator that generates the Dirichlet deformation of the holographic CFT by requiring that the deformed partition function is a Wheeler--DeWitt state and demanding the closure of the algebra generated by the bulk Hamiltonian and momentum constraints. Their deforming operator takes the same form as \eqref{W-partial-flow}. The bulk-boundary match in our energy calculation lends some support to their result.

\subsection{Outlook}\label{sec:outlook}
    There are still open questions which we would like to address in future research. We list some of them here.
    \par As argued in this paper, the equivalence between \TT and Dirichlet deformations does not persist when matter is included, since the Dirichlet operator must include additional contributions from the sources and operators dual to the matter content. However, it is reasonable to expect that some alternative asymptotic MBC, mixing FG metric coefficients as well as scalar sources and operators, might correspond to the Dirichlet deformation, at least classically. Such a reformulation would be useful if one were interested in analyzing the asymptotic symmetry group of this system, as was done for \TT \cite{Guica:2019nzm,Guica:2020uhm,Guica:2022gts,Georgescu:2022iyx}.
    Making this explicit would require repeating the analysis of \cref{sec:3.1} starting from the boundary deformation generated by the appropriate operator $f(J)\left(\mathcal{O}_\TT + \dots\right)$, presumably leading to more complicated solutions for the deformed sources and operators that generalize \eqref{eq:deformedMetricAndStressTensor} and \eqref{defJO}.
    
    In \cref{sec:TTvsDirichlet} we emphasized that both the deforming operator and the flow of the field theory sources are uniquely fixed by the undeformed theory, if we are indeed to reproduce bulk Dirichlet boundary conditions.
    This raises interesting questions as to how such flows differ between distinct microscopic theories which have similar low-energy effective descriptions, and whether such flows can be concatenated.
    Consider for concreteness the following scenario with two examples of AdS/CFT, where CFT$_A$ and CFT$_B$ differ microscopically but have the same CFT dimension and spectrum of single-trace operators below some scale.
    The first question is how much the associated Dirichlet-deformed field theories differ, since they describe similar low-energy physics but are rooted in different UV-complete theories.
    Secondly, one could ask how much of the microscopic content of CFT$_B$ can be recovered from CFT$_A$ by concatenating these flows: one can start from CFT$_A$ and deform it with “Dirichlet deformation $A$” until it describes only a small box in the bulk. (Or only the region just outside of a black hole horizon.) One can then perform “Dirichlet deformation $B$” in the reverse direction all the way back to the AdS boundary.
    The resulting state obtained within this EFT framework could then be compared with CFT$_B$, which contains its exact microscopic description.
    This provides a concrete way to test how much of the microstate information can be recovered by concatenating such Dirichlet flows, an idea which seems to underlie for example the prescription in \cite{Coleman:2021nor} for the microstates of de Sitter space.
    
    \par We would also like to understand more precisely the relation between the holographic interpretation of \TT and its (non-holographic) $2$d-gravitational interpretations \cite{Dubovsky:2017cnj,Dubovsky:2018bmo,Tolley:2019nmm,Callebaut:2019omt}, according to which the deformation is equivalent to coupling the undeformed theory to a dynamical background geometry non-perturbatively.
    In this context, a key role is played by the very same Hubbard--Stratonovich trick described in \cref{sec:3.1} \cite{Cardy:2018sdv}. Thus, we believe that the MBC proposal may provide a holographic embedding to such interpretations. We plan on making this link more explicit, particularly in the metric rather than Chern--Simons formalism. 
    \par In order to match the Dirichlet energy spectrum, we found that it was essential to carefully account for the matter contribution $P$ to the deformed pressure. Curiously, restoring the radius of the cylinder ($\gamma_{ij}dx^idx^j=-dt^2+\mathcal{R}^2 d\varphi^2$) and assuming that the deformed energy may be rewritten as 
    \begin{equation}
    \label{eq:Edimless}
        E_{[\tilde{\rho}_0]}=\tilde{E}_{[\alpha]}/\mathcal{R}
    \end{equation}
    in terms of the dimensionless coupling $\alpha=\tilde{\mu}/\mathcal{R}^2$ would imply that the deformed energy and pressure would be related as $\langle T^{[\tilde{\rho}_0]}_{\varphi\varphi}\rangle=-\mathcal{R}^2\partial_{\mathcal{R}}E_{[\tilde{\rho}_0]}+P_{[\tilde{\rho}_0]}$ (with $P_{[\tilde{\rho_0}]}$ reducing to \cref{P} for $\mathcal{R}=1$). This relation contradicts the standard thermodynamic identity $\langle T_{\varphi\varphi}\rangle=-\mathcal{R}^2\partial_{\mathcal{R}}E$. It is unclear to us whether the assumption \eqref{eq:Edimless} breaks down or whether the generalized relation is, for some reason, invalid.
    \par Finally, we are interested in studying how the addition of bulk matter affects finite-volume holographic dualities under different gravitational boundary conditions -- such as the Neumann boundary condition and the conformal boundary condition, with the recent holographic proposals of \cite{Callebaut:2025thw,Allameh:2025gsa} -- as well as holographic and canonical approaches to quantum gravity beyond AdS, such as the ones developed in \cite{Gorbenko:2018oov,Godet:2024ich,Araujo-Regado:2025elv}.

\section*{Acknowledgments}
    We would like to thank Monica Guica, Richard Myers and Rodolfo Panerai for insightful discussions. BH and MS acknowledge funding by the Deutsche Forschungsgemeinschaft (DFG, German Research Foundation) – Projektnummer 277101999 – TRR 183.

\begin{appendix}
\section{Boundary conditions on the scalar field, general matter counterterms and the alternate quantization}\label{appendixA}

    Consider the matter part of the action for the HMTZ model (we set $l=1$):
    \begin{equation}
			S_m[g,\phi]=-\frac{1}{2\pi G}\int d^3x \sqrt{-g}\left(g^{\alpha\beta}\partial_{\alpha}\phi\partial_{\beta}\phi+2 V_{\nu}(\phi)\right).
    \end{equation}
    Its variation reads
    \begin{equation}
	   	\delta S_m[g,\phi]=\frac{1}{\pi G}\int d^3x \sqrt{-g}\big(\Box \phi -V^{\prime}_{\nu}(\phi)\big)\delta\phi - \frac{1}{\pi G}\int d^2x \sqrt{-h}n^{\mu}\partial_{\mu}\phi \delta \phi +\dots
    \end{equation}
    where $V^{\prime}_{\nu}(\phi)=\partial_{\phi}V_{\nu}(\phi)$ and the dots denote terms proportional to the variation of the bulk metric and of the induced metric at the boundary, which we will omit in what follows unless mentioned explicitly. Note that this matter action already has a well-defined variational principle for a Dirichlet boundary condition on the scalar field consisting in fixing its boundary value, i.e. $\delta \phi \big|_{\partial M} =0$. From the above variation, we can read off the canonical momentum conjugate to the scalar field
    \begin{equation}
       	\pi_{\phi}=\left(\frac{\delta S_m}{\delta \phi}\right)^{\star}=-\frac{\sqrt{-h}}{N \pi G} \dot{\phi},
    \end{equation}
    where we assumed a general ADM decomposition with lapse function $N$, with the dot denoting differentiation with respect to the radial coordinate. Given the precise form of the HMTZ matter potential ($m^2=-3/4$), the scalar wave equation imposes two possible values, $\Delta_+$ or $\Delta_-$, on the conformal dimension $\Delta$ of the dual scalar $\text{CFT}_2$ operator $\mathcal{O}$
    \begin{equation}
       	\Delta=1\pm \frac{1}{2}, \;\;\; \; \; \;\;\; \; \; \;\;\; \; \; \;\;\; \; \; \Delta_+=\frac{3}{2}, \;\;\; \; \; \;\;\; \; \; \;\;\; \; \; \;\;\; \; \; \Delta_-=\frac{1}{2}.
    \end{equation}
    We impose the following asymptotic behaviour on the scalar field\footnote{The relation between the radial FG coordinates used here and in the main text is $\xi=-\frac{1}{2}\log \rho$.}
    \begin{equation}
       	\phi=\alpha\big(x^k\big) \, e^{-\xi/2}+\beta\big(x^k\big) \, e^{-3\xi/2}+\dots
    \end{equation}
    and on the bulk metric
    \begin{align}
       	g_{\mu\nu}dx^{\mu}dx^{\nu}=&d\xi^2 +h_{ij}\big(\xi,x^k\big)dx^idx^j
        \\
        =&d\xi^2+e^{2\xi}\left(\gamma_{ij}^{(0)}\big(x^k\big)+e^{-\xi}\gamma_{ij}^{(1)}\big(x^k\big)+e^{-2\xi}\gamma_{ij}^{(2)}\big(x^k\big)+\dots\right)dx^idx^j,
    \end{align}
    where, here and in the following, dots denote subleading terms in the asymptotic expansions. From now on, we will drop the explicit coordinate dependence of the functions $\alpha$, $\beta$ and $\gamma_{ij}^{(n)}$. Note that $\sqrt{-h}=\sqrt{-\gamma^{(0)}}\left(e^{2\xi}+\sigma e^{\xi}+\dots\right)$, with $\sigma=\frac{1}{2}\gamma^{(0)ij}\gamma^{(1)}_{ij}$. Using $\alpha$ (the leading coefficient in the asymptotic expansion for the scalar field) as the source $J$ for the correlation functions of the dual scalar operator $\mathcal{O}$ is consistent with fixing its conformal dimension to $\Delta_+$ and is equivalent to imposing an asymptotic Dirichlet condition on the scalar field  
    \begin{equation}
    \label{eq:aDirApp}
       	\text{asymptotic Dirichlet b.c.}\left(\Delta=\Delta_+=\frac{3}{2}\right):    \; \; \; \; \; \; \; \; \; \; \;  \delta \alpha=0, \; \; \; \; \; \; \; \; \; \alpha\equiv J.
    \end{equation}
    Using $\beta$ (the subleading coefficient in the asymptotic expansion for the scalar field) as the source $J$ for the correlation functions of the dual scalar operator $\mathcal{O}$ is consistent with fixing its conformal dimension to $\Delta_-$ and is equivalent to imposing an asymptotic Neumann condition on the scalar field  
    \begin{equation}
       	    \text{asymptotic Neumann b.c.}\left(\Delta=\Delta_-=\frac{1}{2}\right):  \; \; \; \; \; \; \; \; \; \delta \beta=0, \; \; \; \; \; \; \; \; \; \beta\equiv J.
    \end{equation}
    Given the unit normal to the asymptotic boundary, with components $n_{\mu}=\delta_{\mu}^{\xi}$ and $n^{\mu}=\delta^{\mu}_{\xi}$, we have the following explicit form of the on-shell variation
    \begin{equation}
       	     \Big(\delta S_m[g,\phi]\Big)^{\star}=\frac{1}{\pi G}\int d^2x \sqrt{-\gamma^{(0)}}\left[\delta \alpha\left(\frac{\alpha}{2}e^{\xi}+\frac{3}{2}\beta+\frac{\alpha \sigma}{2}\right)+\delta \beta \frac{\alpha}{2}\right]+\dots
    \end{equation}
    In order to impose a well-defined asymptotic Dirichlet or Neumann condition on the scalar field, we consider adding boundary terms. We denote these as matter counterterms $S_{ct}^{\phi}[h,\phi]$ because we will also require that they contribute to the renormalization of the gravitational action (and/or change its finite value). These are all the possible covariant combinations of matter boundary terms/counterterms yielding a non-vanishing asymptotic contribution to the action
    \begin{align}
       	    S_{ct}^{\phi}[h,\phi]=&-\frac{1}{8\pi G}\int d^2 x \sqrt{-h}\Big[c_2\phi^2+\tilde{c}_2\phi n^{\mu}\partial_{\mu}\phi+\bar{c}_2\left(n^{\mu}\partial_{\mu}\phi\right)^2+c_4\phi^4+\tilde{c}_4 \phi^3 n^{\mu}\partial_{\mu}\phi \nonumber
       	     \\
       	     &\;\;\;\;\;\;\;\;\;\;\;\;\;\;\;\;\;\;\;\;\;\;\;\;\;\;\;\;\;\;\;\;\;\;\;\;+\bar{c}_4\phi^2\left(n^{\mu}\partial_{\mu}\phi\right)^2+c_4^{\prime}\phi\left(n^{\mu}\partial_{\mu}\phi\right)^3+c_4^{\prime\prime}\left(n^{\mu}\partial_{\mu}\phi\right)^4\Big]
       	    \\
       	    =&-\frac{1}{8\pi G}\int d^2x \sqrt{-\gamma^{(0)}}\Bigg[c_2\left(\alpha^2 e^{\xi}+\alpha^2 \sigma+2\alpha\beta\right)-\tilde{c}_2\left(\frac{\alpha^2}{2}e^{\xi}+\frac{\alpha^2}{2}\sigma+2\alpha\beta\right)\nonumber
       	     \\
       	     &\;\;\;\;\;\;\;\;\;\;\;\;\;+\bar{c}_2\left( \frac{\alpha^2}{4}e^{\xi}+\frac{\sigma \alpha^2}{4}+\frac{3}{2}\alpha\beta\right)+\alpha^4\left(c_4-\frac{\tilde{c}_4}{2}+\frac{\bar{c}}{4}-\frac{c_4^{\prime}}{8}+\frac{c_4^{\prime\prime}}{16}\right)\Bigg]+\dots
    \end{align}
    Clearly, the matter counterterms $S_{ct}^{c_2}$, $S_{ct}^{\tilde{c}_2}$ and $S_{ct}^{\bar{c}_2}$ contribute to leading order relative to the other matter counterterms and, as we will show soon, can cancel the subleading divergence of the unrenormalized action. On the other hand, the leading contribution from the matter counterterms $S_{ct}^{c_4}$, $S_{ct}^{\tilde{c}_4}$, $S_{ct}^{\bar{c}_4}$, $S_{ct}^{c^{\prime}_4}$ and $S_{ct}^{c^{\prime\prime}_4}$ is of order $1$, i.e. they may only shift the finite value of the asymptotic action, hence they are renormalization scheme-dependent. Note that $S_{ct}^{c_2}$ and $S_{ct}^{c_4}$ are the only matter counterterms yielding a well-defined variational principle for a Dirichlet boundary condition on the scalar field consisting in fixing its boundary value, i.e. $\delta \phi \big|_{\partial M} =0$. The on-shell variation of the combined matter action and the above matter counterterms is given by
    \begin{align}
       	     \Big(\delta S_{m}[g,\phi]+\delta S_{ct}^{\phi}[h,\phi]&\Big)^{\star}=-\frac{1}{8\pi G}\int d^2x\sqrt{-\gamma^{(0)}}\Bigg\{\delta\alpha\Bigg[\alpha\left(e^{\xi}+\sigma\right)\left(-4+2c_2-\tilde{c}_2+\frac{\bar{c}_2}{2}\right) \nonumber
       	    \\
       	    &+\beta\left(-12+2c_2-2\tilde{c}_2+\frac{3}{2}\bar{c}_2\right)+4\alpha^3\left(c_4-\frac{\tilde{c}_4}{2}+\frac{\bar{c}_4}{4}-\frac{c_4^{\prime}}{8}+\frac{c_4^{\prime\prime}}{16}\right)\Bigg] \nonumber
       	    \\
       	    &+\delta\beta\Bigg[\alpha\left(-4+2c_2-2\tilde{c}_2+\frac{3}{2}\bar{c}_2\right)\Bigg]\Bigg\}+\dots
    \end{align}
    Clearly, to ensure a well-posed variational principle under the asymptotic Dirichlet condition $\delta \alpha=0$, one must impose the vanishing of the term proportional to $\delta\beta$, which corresponds to setting 
    \begin{equation}
       	    \delta\alpha=0: \;\;\;\;\;\;\; -4+2c_2-2\tilde{c}_2+\frac{3}{2}\bar{c}_2=0.
    \end{equation}
    On the other hand, to have a well-posed variational principle under the asymptotic Neumann condition $\delta \beta=0$, one must impose vanishing of the term proportional to $\delta\alpha$, which corresponds to setting
    \begin{equation}
       	    \delta\beta=0: \;\;\;\;\;\;\;-4+2c_2-\tilde{c}_2+\frac{\bar{c}}{2}=-12+2c_2-2\tilde{c}_2+\frac{3}{2}\bar{c}_2=c_4-\frac{\tilde{c}_4}{2}+\frac{\bar{c}_4}{4}-\frac{c_4^{\prime}}{8}+\frac{c_4^{\prime\prime}}{16}=0.
    \end{equation}
    We are now interested in determining, for each of the above mentioned boundary conditions, which are the appropriate counterterms and thus coefficients yielding a well-posed variational principle and a renormalized action. Consider the unrenormalized action for the HMTZ model supplemented with the GHY term
    \begin{equation} 
       	    S^{(un)}[g,\phi]=\frac{1}{\pi G}\int d^3x \sqrt{-g}\left[\frac{R}{16}-\frac{1}{2}g^{\alpha\beta}\nabla_{\alpha}\phi\nabla_{\beta}\phi-V_{\nu}(\phi)\right]+\frac{1}{8\pi G}\int d^2x \sqrt{-h}K,
    \end{equation}
            which clearly includes the matter action considered above. The Gibbons--Hawking term has the role of ensuring a well-defined Dirichlet condition on the metric, corresponding to fixing the induced metric $h_{ij}$ at the boundary, i.e. $\delta h_{ij}|_{\partial M}=0$ (or, asymptotically, fixing the leading coefficients $\gamma_{ij}^{(0)}$ in its expansion, i.e. $\delta\gamma_{ij}^{(0)}=0$). Considering the HMTZ solutions \cref{ESMbulkmetricFGgauge} and \cref{phiexpansion}, which amounts to fixing
    \begin{equation}
       	    \alpha=\sqrt{B}, \; \; \; \; \; \; \; \; \; \; \beta=\frac{B^{3/2}}{3}, \; \; \; \; \; \; \; \; \; \gamma_{ij}^{(0)}=\eta_{ij}, \; \; \; \; \; \; \; \; \; \gamma_{ij}^{(1)}=-4B\eta_{ij}, \; \; \; \; \; \; \; \; \;\dots
    \end{equation}    
    we computed explicitly
    \begin{equation}
        	S^{(un)\star}[g,\phi]=\frac{1}{8\pi G}\int d^2x \left(e^{2\xi}-2Be^{\xi}\right)+\dots
    \end{equation}
    In order to renormalize the leading $O(e^{2\xi})$ asymptotic divergence while ensuring a well-defined variational principle for the scalar matter field, we must add to the matter counterterms the gravitational counterterm
    \begin{align}
       	    S_{ct}[h,\phi]=&-\frac{1}{8\pi G}\int d^2x \sqrt{-h}+S_{ct}^{\phi}[h,\phi]
            \\
            =&-\frac{1}{8\pi G}\int d^2x \left(e^{2\xi}-4B e^{\xi}+6B^2+\dots\right)+S_{ct}^{\phi}[h,\phi].
    \end{align}
    Thus, the on-shell renormalized action takes the form
    \begin{equation}
         	S^{(ren)\star}[g,\phi]=S^{(un)\star}[g,\phi]+S_{ct}[h,\phi]=\frac{1}{8\pi G}\int d^2x \, e^{\xi}B\left(2-c_2+\frac{\tilde{c}_2}{2}-\frac{\bar{c}_2}{4}\right)+\dots
    \end{equation}
    From this expression, we see that renormalization requires imposing the following relation between the coefficients of the leading matter counterterms
    \begin{equation}
       	    \lim\limits_{\xi\rightarrow\infty}S^{(ren)\star}[g,\phi]=O(1): \; \; \; \; \; \; \; \; \; \; 2-c_2+\frac{\tilde{c}_2}{2}-\frac{\bar{c}_2}{4}=0.
    \end{equation}
    Combining the renormalization condition with the requirement of a well-posed variational principle, we find that for the asymptotic Dirichlet condition (consistent with the conformal dimension of the dual scalar operator being $\Delta=\frac{3}{2}$) we must impose
    \begin{equation}
       	    \delta\alpha=0, \; \; \lim\limits_{\xi\rightarrow\infty}S^{(ren)\star}[g,\phi]=O(1): \; \; \; \; \; \; \; \; \; \; c_2=2+\frac{\tilde{c}_2}{4}=2+\frac{\bar{c}_2}{4}.
    \end{equation}
    For the asymptotic Neumann condition (consistent with the conformal dimension of the dual scalar operator being $\Delta=\frac{1}{2}$) we must have
    \begin{equation}
            \delta\beta=0, \; \; \lim\limits_{\xi\rightarrow\infty}S^{(ren)\star}[g,\phi]=O(1): \nonumber
    \end{equation}
    \begin{equation}
            c_2=\frac{\tilde{c}_2}{4}=\frac{\bar{c}_2}{4}-2 \;\; \wedge \; \; c_4-\frac{\tilde{c}_4}{2}+\frac{\bar{c}_4}{4}-\frac{c_4^{\prime}}{8}+\frac{c_4^{\prime\prime}}{16}=0.
    \end{equation}
    Finally, for a Dirichlet condition valid even for a boundary at finite distance (consistent asymptotically with the conformal dimension of the dual scalar operator being $\Delta=\frac{3}{2}$) we should set
    \begin{equation}
             \delta\phi\Big|_{\partial M}=0, \; \; \lim\limits_{\xi\rightarrow\infty}S^{(ren)\star}[g,\phi]=O(1): \nonumber
    \end{equation}
    \begin{equation}
         	c_2=2 \; \; \wedge \;\; \tilde{c}_2=\bar{c}_2=\tilde{c}_4=\bar{c}_4=c^{\prime}_4=c^{\prime\prime}_4=0.
    \end{equation}
    Notice that, while for the asymptotic Dirichlet boundary condition on the scalar field the requirement of a well-posed variational principle is not necessarily consistent with renormalization of the action, for the asymptotic Neumann condition this is instead the case. Imposing an asymptotic Dirichlet boundary condition on the scalar field ($\Delta=\frac{3}{2}$), we find
    \begin{equation}
            \delta\alpha=0, \; \; \; \; J=\alpha=\sqrt{B}, \; \; \; \;  \lim\limits_{\xi\rightarrow\infty}S^{(ren)\star}[g,\phi]=O(1): \nonumber 
    \end{equation} 
    \begin{equation}
         	S^{(ren)\star}=\frac{B^2}{8\pi G}\int d^2x \left(\frac{2}{3}-c_4+\frac{\tilde{c}_4}{2}-\frac{\bar{c}_4}{4}+\frac{c_4^{\prime}}{8}-\frac{c_4^{\prime\prime}}{16}\right)+\dots
    \end{equation}
    \begin{equation}
       	    \left(\delta S_m[g,\phi] + \delta S_{ct}^{\phi}[h,\phi]\right)^{\star}=\frac{1}{8\pi G}\int d^2x \; \delta{\alpha} \; B^{3/2} \left[\frac{8}{3}-4\left(c_4-\frac{\tilde{c}_4}{2}+\frac{\bar{c}_4}{4}-\frac{c_4^{\prime}}{8}+\frac{c_4^{\prime\prime}}{16}\right)\right]+\dots
    \end{equation}
    \begin{equation}
       	     \langle \mathcal{O}\rangle=\frac{B^{3/2}}{8\pi G}\left[\frac{8}{3}-4\left(c_4-\frac{\tilde{c}_4}{2}+\frac{\bar{c}_4}{4}-\frac{c_4^{\prime}}{8}+\frac{c_4^{\prime\prime}}{16}\right)\right],
    \end{equation}
    \begin{equation}
       	    \langle T_{ij}\rangle =\frac{\gamma_{ij}^{(2)}}{8\pi G}-\frac{B^2}{\pi G}\left[\frac{2}{3}+\frac{1}{8}\left(c_4-\frac{\tilde{c}_4}{2}+\frac{\bar{c}_4}{4}-\frac{c_4^{\prime}}{8}+\frac{c_4^{\prime\prime}}{16}\right)\right]\gamma_{ij}^{(0)},
    \end{equation}
    whereas imposing an asymptotic Neumann boundary condition on the scalar field ($\Delta=\frac{1}{2}$), we obtain
    \begin{equation}
         	\delta\beta=0, \; \; \; \; J=\beta=\frac{B^{3/2}}{3}, \; \; \; \; \lim\limits_{\xi\rightarrow\infty}S^{(ren)\star}[g,\phi]=O(1): \nonumber 
    \end{equation} 
    \begin{equation}
       	    S^{(ren)\star}=-\frac{B^2}{4\pi G}\int d^2x  +\dots
    \end{equation}
    \begin{equation}
         	\left(\delta S_m[g,\phi] + \delta S_{ct}^{\phi}[h,\phi]\right)^{\star}=-\frac{\sqrt{B}}{\pi G}\int d^2x \; \delta{\beta} +\dots
    \end{equation}
    \begin{equation}
         	\langle \mathcal{O} \rangle=-\frac{\sqrt{B}}{\pi G},
    \end{equation}
    \begin{equation}
          	\langle T_{ij}\rangle=\frac{\gamma_{ij}^{(2)}}{8\pi G}-\frac{B^2}{\pi G}\gamma_{ij}^{(0)},
    \end{equation}
    where $\langle \mathcal{O}\rangle =\frac{1}{\sqrt{-\gamma^{(0)}}}\frac{\delta S^{(ren)\star}}{\delta J}$, $\langle T_{ij} \rangle =-\frac{2}{\sqrt{-\gamma^{(0)}}}\frac{\delta S^{(ren)\star}}{\delta\gamma^{(0)ij}}=-\lim\limits_{\xi\rightarrow\infty}\frac{2}{\sqrt{-h}}\frac{\delta S^{(ren)\star}}{\delta h^{ij}}$ and thus $\langle T_{ij}\rangle=-\lim\limits_{\xi\rightarrow\infty}\frac{1}{8\pi G}[K_{ij}-Kh_{ij}+(1+c_2\phi^2+\tilde{c}_2\phi n^{\mu}\partial_{\mu}\phi+\bar{c}_2\left(n^{\mu}\partial_{\mu}\phi\right)^2+c_4\phi^4+\tilde{c}_4 \phi^3 n^{\mu}\partial_{\mu}\phi+\bar{c}_4\phi^2\left(n^{\mu}\partial_{\mu}\phi\right)^2+c_4^{\prime}\phi\left(n^{\mu}\partial_{\mu}\phi\right)^3$ $+c_4^{\prime\prime}\left(n^{\mu}\partial_{\mu}\phi\right)^4)]$. While the asymptotic Dirichlet condition on the scalar leaves some renormalization-scheme dependency, allowing us to change the finite value of the renormalized asymptotic action and one-point functions by tuning the $\{c_4,\tilde{c}_4,\bar{c}_4,c^{\prime}_4,c^{\prime\prime}_4\}$ constants, the asymptotic Neumann boundary condition does not, because the condition $c_4-\frac{\tilde{c}_4}{2}+\frac{\bar{c}_4}{4}-\frac{c_4^{\prime}}{8}+\frac{c_4^{\prime\prime}}{16}=0$ prevents these terms from adding a total non-vanishing contribution to the renormalized action. For an asymptotic Dirichlet boundary condition, choosing the minimal subtraction scheme ($MSS$) amounts to imposing
    \begin{equation}
         	\text{MSS} \; (\delta\alpha=0): \; \; \; \; \; \;\; \; \;\; \; \; c_4-\frac{\tilde{c}_4}{2}+\frac{\bar{c}_4}{4}-\frac{c_4^{\prime}}{8}+\frac{c_4^{\prime\prime}}{16}=\frac{2}{3}, \; \; \; \; \; \; \mathcal{O}=0.
    \end{equation}
    In the main text, we only considered imposing an (asymptotic) Dirichlet condition on the scalar field and hence a dual scalar field of dimension $\Delta=\frac{3}{2}$. We also made the choice $c_2=2$, $\tilde{c}_2=0$, $\bar{c}_2=0$ and $\tilde{c}_4=\bar{c}_4=c^{\prime}_4=c^{\prime\prime}_4=0$ which is indeed consistent with both an asymptotic Dirichlet condition (although not the most general choice in this case) and a Dirichlet condition at finite distance, while simultaneously guaranteeing renormalization of the action. All our main results on the holographic \TT deformation in \cref{sec:TTbar} may be generalized by adding the remaining allowed counterterms to the gravitational action. In particular, this amounts to sending $c_4 \rightarrow C$ in all the expressions for the undeformed and deformed sources and operators, with
    \begin{equation}
			\Delta=\frac{3}{2}: \; \; \; \; \; \; \; \; \; \; \; \; \; \; \; \; C=c_4-\frac{\tilde{c}_4}{2}+\frac{\bar{c}_4}{4}-\frac{c_4^{\prime}}{8}+\frac{c_4^{\prime\prime}}{16},
    \end{equation}
    for an asymptotic Dirichlet boundary condition on the scalar, and
    \begin{equation}
	    	\Delta=\frac{1}{2}: \; \; \; \; \; \; \; \; \; \; \; \; \; \; \; \; C=\frac{8}{3} 
    \end{equation}
    for an asymptotic Neumann boundary condition on the scalar. Therefore, the \TT energy spectrum is recovered through the MBC holographic dual under arbitrary choice of scalar quantization and subtraction scheme.

\end{appendix}

\bibliographystyle{JHEP}
\bibliography{bib}
\end{document}